%% file: main.tex
\documentclass[11pt]{article}

\usepackage{amsmath,amssymb,amsthm}
\usepackage[margin=1in]{geometry}
\usepackage{algorithm,algpseudocode}
\usepackage{bm}
\usepackage[colorlinks=true, citecolor=blue!50!black, linkcolor=blue!50!black, urlcolor=blue!50!black]{hyperref}

\usepackage{soul}
\usepackage{framed}
\usepackage{caption}
\usepackage{subcaption}
\usepackage{tikz}
\usepackage{tikz-cd}
\usepackage{enumerate}
\usepackage{enumitem}
\setitemize{noitemsep,topsep=0pt,parsep=0pt,partopsep=0pt}

\usepackage{bbm}
\newcommand{\eps}{\epsilon}

\usepackage{setspace}
\usepackage{booktabs}
\usepackage[capitalise]{cleveref}

\usepackage{float}

\usetikzlibrary{arrows.meta,decorations.pathreplacing,positioning}

\newcommand{\inparen}[1]{\left( #1 \right) }
\newcommand{\inbrak}[1]{\left[ #1 \right] }

\newcommand{\inset}[1]{\left\{ #1 \right\} }

\newcommand{\suchthat}{\ensuremath{~\middle|~}}
\newcommand{\defined}{\ensuremath{\stackrel{\mathrm{def}}{=}}}

\newcommand{\Oopt}{\ensuremath{h}}

\newcommand{\Var}{\ensuremath{\operatorname{Var}}}
\newcommand{\E}{\ensuremath{\operatorname{E}}}

\newcommand{\ord}[3]{\ensuremath{{#1}_{#3,#2}}}
\newcommand{\Ind}{\mathbbm{1}}
\newcommand{\SR}[1]{\ensuremath{\mathfrak{s}_{#1}}}
\newcommand{\CR}[1]{\ensuremath{\mathfrak{c}_{#1}}}

\newtheorem{lemma}{Lemma}
\newtheorem{theorem}{Theorem}
\newtheorem{corollary}{Corollary}

\newtheorem{definition}{Definition}

\newcounter{protocol}

\newtheorem{example}{Example}

\usepackage{natbib}
 \bibpunct[, ]{(}{)}{,}{a}{}{,}%

\begin{document}
\pagestyle{plain}
\title{Economics of NFTs: The Value of Creator Royalties}

\author{Brett H. Falk, Bin Gu, Gerry Tsoukalas, Niuniu Zhang\footnote{Falk: \href{mailto:fbrett@cis.upenn.edu}{fbrett@cis.upenn.edu}, Gu: \href{mailto:bgu@bu.edu}{bgu@bu.edu}, Tsoukalas: \href{mailto:gerryt@bu.edu}{gerryt@bu.edu}, Zhang: \href{mailto:niuniu.zhang.phd@anderson.ucla.edu}{niuniu.zhang.phd@anderson.ucla.edu}}   }


\date{February 8, 2026}
\maketitle

\bibliographystyle{agsm}

\begin{abstract}
\begin{spacing}{1.}
Non-Fungible Tokens (NFTs) are transforming how content creators, such as artists, price and sell their work. A key feature of NFTs is the inclusion of royalties, which grant creators a share of all future resale proceeds. Although widely used, critics argue that sophisticated speculators, who dominate NFT markets, simply price in royalties upfront, neutralizing their impact. We show this intuition holds only under perfect, frictionless markets. Under more realistic market conditions, royalties enable creators to capitalize on the presence of speculators in at least three ways: They can enable risk sharing (under risk aversion), mitigate information asymmetry (when speculators are better informed), and unlock price discrimination benefits (in multi-unit settings). Moreover, in all three cases, royalties meaningfully expand trade, implying increased transaction volume for platforms. These results offer testable predictions that can guide both empirical research and platform design.



\end{spacing}
\end{abstract}

\section{Introduction}
Artists have long struggled to capture ongoing value as their work appreciates and changes hands. Van Gogh famously died in poverty while his paintings now sell for hundreds of millions. Digital content introduced an additional challenge: perfect copies indistinguishable from originals undermined the scarcity that underpins physical markets. Non-fungible tokens (NFTs) promise to address both issues at once: establishing verifiable ownership while embedding royalties directly into digital assets, enabling creators to capture value from every resale through programmable, automated payments and without traditional gatekeepers.

This vision is already taking shape. By 2022, over \$1.8 billion in royalty payments had flowed back to creators on Ethereum alone \citep{GD22}, and over \$50 billion worth of NFTs have been sold since inception \citep{cointelegraph_nft_2024}. For many creators, royalty revenues have dwarfed those from the initial sale \citep{P22}. The rates themselves are substantial: \citet{Tunc2024NFTRoyalty} document that rates requested by creators average around 10\% on Rarible; our own examination of rates actually enforced across three major exchanges---OpenSea, LooksRare, and X2Y2---yields an average of approximately 7.35\% (see Appendix~\ref{sec:empirical} for details).

But as promising as the technology appears, the long-term economic viability of creator royalties faces two distinct criticisms. The first is mechanical: pricing-in. Because royalties represent a cost that must be paid on every future resale, rational buyers will incorporate this expected cost into their bidding strategies, reducing the price they are willing to pay upfront, thereby neutralizing the royalty's effect entirely.\footnote{For example, this pricing-in argument is put forward by Haseeb Qureshi on The Chopping Block \url{https://youtu.be/oRWvJlhYnLc?t=3116}} Under this logic, a creator who raises her royalty rate simply receives less from the initial sale, leaving total revenue unchanged. Consistent with this view, higher royalty rates correlate with systematically lower initial sale prices \citep{Tunc2024NFTRoyalty}.

The second criticism is structural: NFT markets are shaped not only by casual collectors but by sophisticated market players   \citep{Reuters2021NFT, digitalveblen, niedermayer2024detecting, Tunc2024NFTRoyalty, falk2025can} (which we will refer to as ``speculators'' going forward) who mediate between creators and end-buyers. In idealized markets, these intermediaries would be mere pass-throughs, unable to capture any surplus beyond what the creator could extract directly. But real markets are characterized by well-known imperfections: risk-bearing capacity varies across participants \citep{pratt1964, kahneman1979}, information is unevenly distributed \citep{Akerlof1970, Grossman1980}, and some agents can tailor prices to heterogeneous buyers while others cannot \citep{Mussa1978, Varian1985}. These imperfections allow intermediaries to do more than pass through value---they enable them to capture it.

NFT speculators exploit each of these channels. They diversify across collections to absorb idiosyncratic project risk and invest in trading infrastructure that delivers faster demand signals. And unlike creators locked into an ex ante mint price, they can use block-level trading strategies \citep{daian2020flash} to adjust execution timing and markups based on transaction-level demand signals and competitive pressure. Together, these advantages give speculators considerable leverage over creators in both primary and secondary NFT markets.

These criticisms stand in tension with the market evidence: if speculators can price in royalties, and hold such pronounced advantages over creators, why do royalties remain so widely used and economically significant? This tension motivates the following research questions:
\begin{enumerate}[label=\textbf{\Roman*.},nosep]
\item Can sophisticated speculators fully price in creator royalties, neutralizing their impact?
\item Do creator royalties generate economic value for creators, and if so, how?
\item How do royalties affect the market participation incentives of creators and speculators?
\end{enumerate}

To address these questions, we develop a stylized model where a creator mints one or more NFTs for sale. In line with how these sales are conducted in practice, the creator must choose a fixed mint price and has the option of attaching a percentage royalty.\footnote{Alternative selling mechanisms like auctions have been gaining popularity, but fixed-price selling remains the more dominant mechanism due to its simplicity and immediacy \citep{opensea_sell_guide, coingecko_opensea_guide}.} For royalties to take effect, we require at least a two-period model in which a future trade can trigger the royalty payment. Sophisticated speculators are present who seek to purchase NFTs from the creator before other buyers, with the hope of reselling at a later date if prices appreciate \citep{Reuters2021NFT}. Speculators are strategic: they make purchase decisions based on expected future returns, which include any royalty obligations. End-buyer valuations are uncertain at the time of the initial sale, creating potential upside but also downside risk for both creators and speculators.
We first analyze an idealized benchmark in which the creator and speculator have the same information and are both risk neutral. This establishes whether royalties can add value on their own, before any market imperfections are introduced. We then examine how each of the three imperfections described above --- differential risk tolerance, information asymmetry, and pricing power  --- interacts with creator royalties.

We find that in the idealized benchmark, the pricing-in critique is vindicated: royalties are rendered economically irrelevant in the presence of speculators, who fully offset any royalty obligation by reducing what they pay upfront.

But this irrelevance breaks down once we introduce the market imperfections described above, and it does so in a surprising direction. Rather than compounding speculators' advantages over creators, royalties allow creators to piggyback on them.

When creators and speculators differ in their tolerance for risk, royalties function as a risk-sharing mechanism that increases creator utility and facilitates trade. Without royalties, the speculator bears the entire variance of the uncertain resale price and demands a steep upfront discount to compensate. Even though creators are the more risk-averse party, accepting some of this risk exposure through royalties can leave them better off: the speculator, relieved of part of the risk burden, requires a smaller discount, and the creator gains a direct claim on aftermarket upside that more than compensates for the additional risk absorbed.

When speculators possess superior information, royalties allow creators to retain upside from future price appreciation and again expand the trade region. The intuition is that a creator who cannot observe market conditions can still participate in favorable outcomes by tying her compensation to the resale price. Royalties are particularly well-suited to this setting because creators commit to a fixed mint price at the smart contract level, limiting their ability to adjust prices conditional on observed behavior.

And when creators mint multi-unit collections, royalties enable price discrimination across buyers with heterogeneous valuations, once more expanding trade. Here the creator faces a different constraint: she must set a single mint price for all units, but the speculator can resell individual items at differentiated prices. Royalties allow the creator to share in this differentiated revenue.

In all three cases, these benefits persist even when fully rational buyers price royalty costs in at the time of purchase.

We probe the robustness of these findings along several dimensions. In the appendix, we extend the model to incorporate speculator outside options, which yields more realistic equilibrium royalty rates; creator overconfidence; partial surplus extraction by speculators; and liquidity effects on end-buyers, which reveals that under some market conditions the optimal royalty rate is zero. Our core results survive each extension.

Our results carry practical implications for the ongoing royalty debate. NFT royalty enforcement has been contentious: early standards such as EIP-2981 allowed creators to specify a royalty rate but left compliance at the discretion of individual platforms. Marketplace stances have ranged from full enforcement to outright bypass. Yet as documented above, substantial royalty flows materialized even under this discretionary regime. Recent standards such as ERC-721C on Ethereum and Metaplex's Programmable NFT on Solana have made meaningful progress by embedding enforcement logic at the smart contract or protocol level \citep{opensea_erc721c_2024, magiceden_mip1_2024}. As enforcement becomes increasingly feasible, the question shifts from whether royalties \emph{can be} collected to whether they \emph{should be}. Our findings  suggest an affirmative answer (at least from the creator's perspective): royalties are not merely a tax on secondary market activity but a mechanism through which creators can harness the very advantages that sophisticated speculators hold over them. If royalties generate value through risk-sharing, information revelation, and price discrimination, then enforcing them is not merely honoring creator preferences but facilitating welfare-improving outcomes. We discuss these implications for platform governance in greater detail in Section~\ref{sec:discussion}.

These findings extend beyond NFTs to any setting where creators sell rights to assets that sophisticated intermediaries may later trade — including music licensing, ticket resale, and real estate presales. More broadly, our analysis suggests that creator royalties play a previously underappreciated role in the economics of markets characterized by both creative production and speculative activity. 

\section{Literature Review}\label{sec:lit}

Our analysis draws on and contributes to three distinct literatures: speculation in markets, royalty contracts, and blockchain-based NFTs.

The literature on speculation reflects a fundamental tension. \citet{keynes1936general} characterizes speculation as a destabilizing force that transforms markets into casinos, while \citet{Grossman1980} argue it serves an essential price discovery function when information acquisition is costly. Recent work has examined how speculators affect market outcomes across different contexts. \citet{Courty2003} demonstrates that ticket scalpers leave event promoters worse off under demand uncertainty, while \citet{optimalpricingwithspeculators} shows how speculators extract surplus when sellers face pricing constraints. In real estate, \citet{defusco2022speculative} find that informed speculators amplify trade volume when sellers face timing constraints, and in cryptocurrency markets, \citet{daian2020flash} document how algorithmic arbitrageurs rapidly exploit pricing inefficiencies. Building on this literature, our model treats speculators as strategic intermediaries who exploit market imperfections to gain advantages over creators. Our contribution is showing that royalty contracts can transform what is typically viewed as surplus extraction into a tool for creator benefit.

Royalties, often structured through licensing agreements, have been extensively analyzed in the economics literature. Early seminal work by \citet{Arrow1962} establishes intellectual property rights, often implemented through royalties, as critical incentives for innovation. \citet{Gallini1984} demonstrates that incumbents can use royalty-based licensing strategically to deter entry by sharing market profits, thereby reducing competitive pressures. On the other hand, several works argue that sophisticated buyers pose genuine threats to licensors' ability to capture surplus from their intellectual property. \citet{Kankanhalli2018} show that larger licensees negotiate significantly lower royalty rates than smaller competitors, and licensors often redact unfavorable payment terms in public filings. Our paper shows that the standard finding --- that sophisticated buyers erode sellers' surplus --- can reverse when royalty contracts allow sellers to participate in speculator-generated gains. We find that, at least within the context of NFTs, royalties are most effective for sellers when buyers have informational and trading advantages.

Several recent works empirically analyze NFT markets. Closest to our study, \citet{Tunc2024NFTRoyalty} use transaction data from Rarible to examine how royalties affect prices, liquidity, and creator welfare. On prices, our papers agree: higher royalties lead to lower primary market prices. \citet{Tunc2024NFTRoyalty} attribute this to a delayed gratification effect; our model explains this through the equilibrium trade-off between immediate revenue and the expected royalty stream. On welfare and liquidity, the papers are complementary because they examine different margins. Our model concerns creators' \emph{ex ante} participation decision and shows that royalties expand the set of participating creators by allowing them to extract surplus from speculators. \citet{Tunc2024NFTRoyalty} study \emph{ex post} performance of NFTs already listed, finding a negative reduced-form correlation between royalty rates and total creator revenue.%
\footnote{Their log-log specification is undefined at zero royalties and imposes monotonicity, so the estimated elasticity is identified only from variation among positive rates and cannot speak to whether creators benefit from the \emph{option} to set royalties. Selection may also contribute: creators with weaker commercial prospects may set higher rates, generating a negative association even absent a causal effect.} 
However, their data condition on participation and so cannot capture counterfactual creators who would not have listed without royalties. Critically, \citet{Tunc2024NFTRoyalty} show that the negative relationship is not uniform across creators, and trace the negative correlation to overconfident creators who aggressively lower primary prices but fail to recoup through royalties. Our framework can account for this directly: in Appendix~\ref{app:overconfidence}, we show that an overconfident creator sets higher royalty rates and earns strictly less total revenue, generating exactly this pattern. \citet{Tunc2024NFTRoyalty} also find that royalties reduce \emph{ex post} market liquidity. In Appendix~\ref{app:liquidity}, we extend the model to allow for liquidity effects on end-buyers, and our main welfare results remain robust, with the added nuance that for some market conditions the optimal royalty rate is zero.

Related empirical work documents speculative behavior broadly consistent with our modeling assumptions: \citet{falk2025can} find evidence of significant market manipulation in NFT markets, while \citet{EconsofNFTs} and \citet{FundamentalvalueofNFTs} analyze the determinants of NFT returns and valuation. On the conceptual side, several papers discuss the potential benefits of creator royalties \citep{NewForms, fractionalequity}, though without formal economic modeling, and \citet{digitalveblen} analyze NFTs as digital Veblen goods. Our work also connects to the broader literature on the economics of token systems, which primarily focuses on fungible tokens \citep{cong2021tokenomics, benhaimscaling, gan2021initial, gan2021infinity, saleh2021blockchain, tsoukalas2020token}.

\section{General Model}
\label{sec:model}

In this section, we describe our general model of NFT sales that underpins the specialized models in subsequent sections.  There are three periods and three types of agents.  

At time 0, a ``creator'' produces an item (or multiple items) that she wishes to sell by minting a corresponding NFT. Production and minting costs  are denoted $c\geq0$. As mentioned in the introduction, she sets her fixed NFT mint price, $p_0\geq0$, and possibly a royalty rate $r \in [0,1]$ which entitles her to receive (in addition to $p_0$) an $r$-fraction of future sales proceeds, if any. 

At time $1$, a ``speculator'' can choose to purchase the NFT at $p_0$. He does so only if he expects to be able to resell it later and break even with respect to his outside option, initially normalized to zero. If trade does not occur, the speculator exits the market. 

The speculator should be thought of as an arbitrageur or scalper that can have some advantages over the creator. This division of labor between creators and speculators is standard in the literature on speculation (see, e.g., \citealt{Courty2003}; \citealt{karp2005when}; \citealt{optimalpricingwithspeculators}; \citealt{gan2021initial}) and reflects empirical differences in costs, skills, and objectives: creators specialize in producing content, while speculators invest in trading infrastructure such as MEV extraction and bot networks \citep{daian2020flash, niedermayer2024detecting}. We model these advantages in three ways: in Section~\ref{sec:riskaversion}, the speculator has higher risk tolerance; in Section~\ref{sec:information_asymmetry} he has informational advantages; in Section~\ref{sec:price_discrimination} he has price discrimination advantages.

At time $2$, an end-buyer appears who is willing to purchase the NFT at a random valuation, $V$, with mean $\mu$, standard deviation $\sigma$, PDF $f$, and CDF $F$ (discrete distributions are also admissible). The distribution of $V$ is known to both creator and speculator, however, the speculator has the following advantage: If he owns the NFT at time 2, he does not have to commit to selling it at the creator's mint price $p_0$. Instead, the speculator can sell it at the realization of $V$---this is a stylized assumption to capture the speculator's superior ability to extract buyer surplus \citep{defusco2022speculative, Courty2003, karp2005when}. We relax this assumption in Appendix~\ref{app:partial_extraction}, showing that qualitative results hold under partial surplus extraction as well.

If the speculator transacts with the end-buyer at time 2, a percentage $r$ of the sales proceeds is remitted to the creator. Conversely, if the speculator chooses not to purchase the NFT in period 1, then the creator's NFT is still available to the end-buyer at time 2 (at the original mint price $p_0$). In this case, trade occurs only if the realization of $V \ge p_0$;  that is, the end-buyer observes their valuation at the time of purchase and trades only if it exceeds the mint price. This setup is meant to capture the dominant model of NFT sales in practice, where creators usually sell their NFTs at a fixed mint price, whereas resellers (speculators) operate primarily in the NFT aftermarket searching for the best returns and arbitrage opportunities. 

Two modeling notes deserve comment. First, the sequential structure---where the speculator arrives before the end-buyer---follows \citet{gan2021initial} and serves as a tractable proxy for the well-documented speed advantages of sophisticated actors in blockchain markets \citep{daian2020flash}. Simultaneous arrivals as in \citet{optimalpricingwithspeculators} would introduce additional strategic considerations. Second, following standard practice in mechanism design (see, e.g., \citealt{krishna2010auction}), we assume throughout the paper that if speculators are indifferent between purchasing NFTs and abstaining, they choose to purchase.

Using this model as a base, we consider three specializations:

\textbf{Risk Aversion:} In Section~\ref{sec:riskaversion}, we consider a situation where both the creator and the speculator are risk averse, specifically the creator and the speculator have a mean-variance utility.

\textbf{Information Asymmetry:} In Section~\ref{sec:information_asymmetry}, we consider the situation where the speculator can be better informed regarding the future value of the NFT. Specifically, he learns the realization of $V$ before he needs to make a decision on purchasing the NFT at time $1$.  This now gives the speculator an informational advantage over the creator, who only knows the \emph{distribution} of $V$ at the time she chooses $p_0$. 

\textbf{Buyer Heterogeneity:} In Sections~\ref{sec:riskaversion} and \ref{sec:information_asymmetry}, we assume the creator has a single NFT to sell.  In Section~\ref{sec:price_discrimination}, we consider the situation where the creator has more than one NFT to sell to buyers with heterogeneous valuations.

\begin{figure}[ht]
\centering
\begin{tikzpicture}[
  font=\small,
  >=Latex,
  timeline/.style={line width=0.9pt},
  tick/.style={line width=0.9pt},
  topnote/.style={align=center, font=\small},
  botnote/.style={align=center, font=\small},
]

\def\xL{0}
\def\xR{13.6}
\def\xTzero{1.4}
\def\xTone{6.6}
\def\xTtwo{11.2}
\def\yLine{0}

\draw[timeline,->] (\xL,\yLine) -- (\xR,\yLine) node[right] {Time};

\draw[tick] (\xTzero,\yLine+0.12) -- (\xTzero,\yLine-0.12);
\draw[tick] (\xTone,\yLine+0.12)  -- (\xTone,\yLine-0.12);
\draw[tick] (\xTtwo,\yLine+0.12)  -- (\xTtwo,\yLine-0.12);

\node at (\xTzero,\yLine-0.45) {$t=0$};
\node at (\xTone,\yLine-0.45)  {$t=1$};
\node at (\xTtwo,\yLine-0.45)  {$t=2$};

\def\yText{1.05}   
\def\yArrow{0.55}  

\draw[->] (\xTzero,\yArrow) -- (\xTzero,\yLine+0.18);
\node[topnote] at (\xTzero,\yText) {Creator chooses\\$(p_0,r)$};

\draw[->] (\xTone,\yArrow) -- (\xTone,\yLine+0.18);
\node[topnote] at (\xTone,\yText) {Speculator\\buy/abstain};

\draw[->] (\xTtwo,\yArrow) -- (\xTtwo,\yLine+0.18);
\node[topnote] at (\xTtwo,\yText) {End-buyer(s)\\buy/abstain};

\node[botnote] at (\xTzero,-1.05) {NFT Minted};
\node[botnote] at (\xTone,-1.05) {Primary sale\\(if any)};
\node[botnote] at (\xTtwo,-1.05) {Secondary sale\\(if any)};

\end{tikzpicture}
\caption{\footnotesize Sequence of events in the general model.}
\label{fig:sequence_of_events_nft}
\end{figure}

\section{Benchmark: When are Royalties Not Needed?}\label{sec:benchmark}

In this section, we analyze the significance of royalties in the bare-bones model which will serve as a benchmark, setting aside risk aversion, information asymmetry, and multi-unit collections. Our focus here is on understanding three things: (i) what does trade look like when the creator is directly selling to the end-buyer? (ii) under what conditions can speculators add value as middle-men between creator and end-buyer? (iii) when are royalties NOT needed?

\subsection{Direct Trade with an End-Buyer (No Speculator Present)}\label{sec:benchdirect}

For now, consider the speculator is absent entirely and hence time 1 has no actions. By default, this implies royalties cannot play a role here (as mentioned, one requires at least two periods of trade to not trivially rule them out). Instead, the creator mints a NFT at time 0, hoping to sell it directly to an end-buyer with random valuation $V$ at time 2. 

As discussed, the creator has to set the mint price $p_0$ before the realization of $V$ is known. One important implication is that for many common distributions of $V$, the optimal mint price will not be $p_0^* = \E[V]$, and the creator will not able to fully extract end-buyer willingness-to-pay, $\E[V]$. To see this, note that if $p_0 \leq V$, then the buyer buys the item, and the creator earns $p_0$.  On the other hand, if $V < p_0$, there is no trade and the creator earns nothing.  Let $\mathbbm{1}_{ V \ge p_0 }$ be the corresponding indicator function. The creator's expected profits are given by $p_0 \cdot \E[ \mathbbm{1}_{ V \ge p_0 } ] - c$, which can be written as $p_0 \cdot \Pr [ V \ge p_0 ] - c.$ Her optimization problem can be written as
\begin{equation}\label{eq:objbare}
    \max_{p_0} \quad p_0 \cdot \Pr [ V \ge p_0 ] - c.
\end{equation}
Without loss of generality, we can set $c=0$ here given it is an additive constant. For most common distributions, we have the following result:
\begin{lemma}\label{lem:optpbase}
When the creator is directly selling to an end-buyer, and assuming unimodality of the objective in \eqref{eq:objbare}, the optimal mint price, $p_0^*$, is given by the solution to the following first-order condition:
\begin{align}
    p_0 f(p_0) = 1-F(p_0).
\end{align}
\end{lemma}
Depending on the distribution of $V$, the above may NOT have a closed-form solution. Even so, we can characterize the creator's optimal profits without explicitly solving for the optimal mint price. By Markov's inequality (also known as the first Chebyshev inequality; see \cite{grimmett2001probability}), regardless of the distribution of $V$ and of the creator's choice of $p_0$, the revenues satisfy
\begin{equation}\label{eq:markov}
    p_0 \cdot \Pr [ V \ge p_0 ] \le \E[V], \quad \forall \ p_0 \in [0,\infty),
\end{equation}
with the inequality being strict whenever $V$ is non-degenerate---that is, when it is not deterministic and has non-zero variance (see Appendix~\ref{app:6pages-nondegen} for the formal definition).

Taken together, these arguments imply the following corollary:
\begin{corollary}[Speculator Absent]\label{cor:nono}
In the absence of a speculator, and without the possibility of royalties, the creator's optimal revenues are given by $\max_{p_0} p_0 \Pr[V\geq p_0]$, and Markov's inequality \eqref{eq:markov} implies she cannot fully extract expected end-buyer willingness-to-pay, $\E[V]$.
\end{corollary}

\subsection{How Speculators can Improve Trade Outcomes (Absent Royalties)}\label{sec:spec_improve_trade}

As discussed in the introduction, speculation is  rampant in NFT markets, and in this section we examine some of the implications of having a speculator present who can serve as a trade intermediary between creator and end-buyer. Royalties are still ruled out for the moment.

Suppose there exists a speculator arriving at time $1$, as described in the general model in Section~\ref{sec:model}. For the purposes of the benchmark model, the speculator and creator have the same information at the time the NFT purchasing decision needs to be made. However, the speculator has the advantage of operating in the aftermarket at time $2$. There, he can sell the NFT to an end-buyer at the end-buyer's realization of $V$. This implies that in expectation, the speculator extracts $\E[V]$ from the buyer. The full-extraction assumption is a simplification relaxed in Appendix~\ref{app:partial_extraction}.

Denote the speculator's sale price at time 2, $p_2$. At time 1, the speculator is making his purchasing decision from the creator, anticipating that he can earn revenues amounting to $\E[V]$ from the end-buyer, at time 2. The speculator's optimal strategy is to set his own price at time 2, to match the realization of $V$, $p_2 = v$. Thus, Markov's inequality is no longer relevant here.

In turn, the creator can set her own price at time 0 anticipating the speculator's optimal strategy. In particular, the creator can set the mint price $p_0 = \E[V]$, knowing that it is still rational for the speculator to purchase at this price (and break even with respect to his outside option). Compared to the previous Section~\ref{sec:benchdirect}, this can allow the creator to extract higher revenues, and also ensures that the creator will always prefer to sell to the speculator at time 1, rather than wait to sell to the end-buyer at time 2. The main takeaway is formalized in the corollary below.
\begin{corollary}[Speculator Present]\label{cor:noyes}
In the presence of an equally-informed speculator, without the possibility of royalties, the creator can  ``piggyback'' on the speculator's access to the aftermarket to better capture end-buyer willingness-to-pay, obtaining excess profits (compared to the case without the speculator described in Corollary~\ref{cor:nono}) of:
$$\Delta = \E[V] - \max_{p_0} \ \Big ( p_0 \cdot \Pr[ V \ge p_0] \Big )  > 0.$$
\end{corollary}
\noindent The inequality is strict if $V$ is non-degenerate (see \cref{lem:markovloose} in Appendix~\ref{app:6pages-nondegen} for details).

\subsection{When are Royalties NOT Needed?}

Section~\ref{sec:benchdirect} showed that a creator selling directly to an end-buyer cannot fully extract $\E[V]$ because she must commit to a mint price before uncertainty resolves. Section~\ref{sec:spec_improve_trade} showed that a speculator solves this problem: the creator can set $p_0 = \E[V]$ and capture the full surplus, without any royalties. Can royalties do anything more?

Under the current assumptions, no. The logic parallels the Modigliani-Miller theorem in corporate finance \citep{modigliani1958cost}: just as a firm's total value is determined by its underlying cash flows rather than how claims are packaged, the surplus here is pinned down by the speculator's expected aftermarket revenue, $\mu$. The creator is choosing how to package her claim to this surplus---upfront through the mint price, or contingently through royalties---but under risk neutrality and symmetric information, the speculator adjusts his upfront bid to offset any royalty obligation, and repackaging cannot change the size of the pie. Formally:

\begin{theorem}[When are royalties NOT needed?]
    \label{thm:riskneutral}
    When selling a single NFT, if both the creator and speculator have the same information and are risk neutral, then {\bf creator royalties play no effective role}. In particular, any $\{p_0^*, r^*\}$ satisfying $\{ p_0^* = \mu (1-r^*), p_0^* \geq 0, r^* \in [0,1]\}$, maximize the creator's expected profits, which are given by $\mu - c$.
\end{theorem}

All proofs are provided in the appendix. The condition $p_0^* = \mu(1-r^*)$ makes the mechanism transparent: every dollar shifted from mint price to royalties is offset one-for-one by a reduced upfront payment.

But just as Modigliani-Miller irrelevance motivates the study of which imperfections make capital structure matter, Theorem~\ref{thm:riskneutral} raises the central question of this paper: \emph{when does the creator's revenue structure matter?} Mint prices are set \textit{before} uncertainty resolves; royalties are paid \textit{after}, contingent on realized prices. In the benchmark, this timing distinction is irrelevant because the creator can perfectly anticipate aftermarket revenues, the same feature that allowed her to set $p_0 = \E[V]$ in Section~\ref{sec:spec_improve_trade}. But when market imperfections break this ability, the creator can no longer extract surplus through the mint price alone, and the form of the claim begins to matter. The next three sections examine when and how.

\section{Royalties as a Risk-Sharing Mechanism}
\label{sec:riskaversion}

The benchmark result in Theorem~\ref{thm:riskneutral} relies on both parties being risk neutral. In practice, however, creators and
speculators face fundamentally different risk exposures. A creator's livelihood is typically concentrated in a small number of works, making them sensitive to the uncertainty of any single project's commercial outcome. Speculators, by contrast, can diversify across many uncorrelated positions, effectively behaving as if they are closer to risk neutral. This asymmetry in risk bearing is a natural market imperfection: it does not arise from informational problems or transaction costs, but from the structural difference between producing a unique asset and trading a portfolio of them. We now examine how royalties interact with this asymmetry.

{\bf Modifications to base model:} 
Using the notation from Section~\ref{sec:model}, conditional on the speculator owning the NFT at time $2$ and selling it to the end-buyer, profits at time 2 are given by:
\begin{align*}
    v_c(p_0,r) &= p_0 + r p_2  - c \quad \mbox{(Creator profit)}\\
    v_s(p_0,r) &= (1-r)p_2 - p_0 \quad \mbox{(Speculator profit).} 
\end{align*}
We consider a mean-variance utility over profits with risk parameters $\eta_c$ for the creator and $\eta_s$ for the speculator (alternative utility functions can be specified without affecting the qualitative results). Furthermore, as previously discussed, the speculator's optimal strategy at time 2 is simply $p_2 = v$, which extracts the maximum willingness-to-pay from the end-buyer, and ensures trade occurs; adding risk aversion does not alter this result given the speculator's pricing decision is made after uncertainty realization. With this, agent utilities before time 2 can be written as
\begin{align*}
    u_c(p_0,r) &= \E[v_c(p_0,r)] - \eta_c \Var[v_c(p_0,r)] \quad \mbox{(Creator utility)}\\
    u_s(p_0,r) &= \E[v_s(p_0,r)] - \eta_s \Var[v_s(p_0,r)] \quad \mbox{(Speculator utility),} 
\end{align*}
where $\E[v_c(p_0,r)] = p_0 + r \mu - c$, $\E[v_s(p_0,r)] = (1-r)\mu - p_0$, $\Var[v_c(p_0,r)] = r^2\sigma^2$, $\Var[v_s(p_0,r)] = (1-r)^2 \sigma^2$.
With these definitions, the creator's objective is to maximize $u_c$ by choosing the sale price, $p_0$, and creator royalty, $r$, subject to the speculator's participation constraint $u_s \geq 0$.  Formally:
\begin{align}
\max_{p_0, r} &\quad \E\left [ v_c(p_0,r) \right] - \eta_c \Var[v_c(p_0,r)] \label{eqn:creator_utility}\\
& \mbox{subject to:} \quad u_s(p_0, r) \geq 0,
\label{eqn:speculator_profits}
\end{align}
with $u_s(p_0, r) = (1-r)\mu -p_0 -\eta_s (1-r)^2\sigma^2$. The solution is provided in Lemma~\ref{lem:risk}.

\begin{lemma}[Optimal mint price, royalty, and utility]\label{lem:risk}
    Under risk aversion, the creator's optimal NFT mint price, royalties and utility are given by
        \begin{align}
        p_0^* = \frac{\eta_c}{(\eta_c + \eta_s)^2} \Big ( (\eta_c + \eta_s) \mu - \eta_c \eta_s \sigma^2 \Big ), \quad
        	r^*= \frac{\eta_s}{\eta_s + \eta_c}, \quad 
        	u_c^* = \mu - c - \inparen{\frac{\eta_c \eta_s}{\eta_c + \eta_s}} \sigma^2. \label{eqn:optr}
        \end{align}      
        Without royalties, the mint price and utility are given by $p_{0,r=0}^* = \mu - \eta_s \sigma^2,$ and $u_{c,r=0}^* = \mu - c  - \eta_s \sigma^2.$    
\end{lemma}
Lemma~\ref{lem:risk} shows that optimal royalty rates can be strictly positive in the presence of risk aversion. A natural question that follows is how much utility they add to creators. Comparing $u_c^*$ to $u_{c, r=0}^*$ in Lemma~\ref{lem:risk} brings us to the formal result:
\begin{theorem}[Risk Aversion] \label{thm:royalhigh}
    Under risk aversion, creator royalties increase creator utility by 
    $$ \Delta = \sigma^2 \frac{\eta_s^2}{\eta_s+\eta_c}.$$ Further, $\Delta$ increases: (i) as future price uncertainty, $\sigma$, increases; (ii) as the speculator becomes more risk averse (i.e., as $\eta_s$ increases); (iii) as the creator becomes less risk averse (i.e., as $\eta_c$ decreases).
\end{theorem}

\noindent \textbf{Economic Interpretation:} These results reveal that royalties function as a risk-sharing mechanism. Without royalties, the speculator bears the entire variance of the resale price, and demands a steep discount from the creator to compensate. Royalties allow the creator to absorb a share of this variance: because royalty income is contingent on the uncertain resale price, a positive royalty rate shifts some of the speculator's risk exposure back to the creator. This relieves the speculator, who in turn requires a smaller discount, and the creator accepts the exposure willingly because the expected royalty income more than compensates for the risk absorbed. The value of this arrangement depends on three key factors:

First, greater price uncertainty ($\sigma$) makes risk-sharing more valuable, as there is simply more risk to be allocated between the parties. Second, the more risk-averse the speculator ($\eta_s$ higher), the more surplus they are willing to concede to offload variance, improving the terms for the creator. Third, the less risk-averse the creator ($\eta_c$ lower), the more cheaply they can absorb the speculator's risk, increasing the net gains from the arrangement.

To illustrate the risk-sharing mechanism more explicitly, consider the case where both parties exhibit identical risk preferences ($\eta_s=\eta_c=\eta$). The optimal royalty rate then simplifies to $r^* = 1/2$, and the speculator's utility goes to zero, while the creator's utility gain (over the case without royalties) is $\Delta = \sigma^2/2$. This implies that the creator takes on half of the resale price volatility $\sigma^2$, relieving the speculator of that exposure.

{\bf Calibration to Data:} As a hypothetical exercise, we can compute  the relative risk aversion levels implied by  the average royalty rates observed in the empirical section (7.35\%). Assuming creators are on average $x$-times more risk-averse than speculators, we can define $\eta_c = x\cdot \eta_s$, with $x>1$, giving:
\[
r^* = \frac{\eta_s}{\eta_s + x \cdot \eta_s} = \frac{1}{1+x} = 7.35\%. 
\]
Solving gives $x \approx 12.6$: creators would need to be roughly one order of magnitude more risk-averse than speculators for the royalty rates they rationally set in equilibrium to match those observed in practice. While this calibration is stylized and other forces also shape observed rates, the implied differential is consistent with the risk asymmetry discussed above.

\subsubsection*{Royalties Expand Trade Under Risk Aversion}
A secondary benefit of royalties when both creator and speculator are risk averse, is that they can increase the propensity of trade, that is, they expand the space of parameters in which trade occurs between creator and speculator. To see this, note that the results in Lemma~\ref{lem:risk} require the extra condition $u_c^* \geq 0$, otherwise there is no trade. Comparing $u_c^* \geq 0$ to $u_{c,r=0}^* \geq 0$, the former is easier to satisfy. As an example, let's assume $\mu=.5,\sigma=1, c=0,\eta_s=\eta_c=.5$. Then $u_{c}^* = 1/4$ and $u_{c,r=0}^* = 0$, and thus trade would only occur if royalties are allowed (and enforced). More formally, let $\mathbbm{T}$ be the set of parameters for which trade would occur if royalties were allowed, but would not occur in their absence.
\begin{corollary}[Royalties expand the trade region]\label{cor:TRrisk}
Under risk aversion, royalties increase the region of trade, that is, the set $\mathbbm{T} = \{\mu, c, \eta_s, \eta_c, \sigma \ | \ u_c^* \geq 0 \ \cap \ u_{c,r=0}^* < 0  \ \cap 0 \ \leq c \leq \mu  \ \cap \ \eta_c \ge \eta_s \ge 0\}$ is non-empty.
\end{corollary}

\begin{figure}[ht]
    \centering
    \includegraphics[width=250pt]{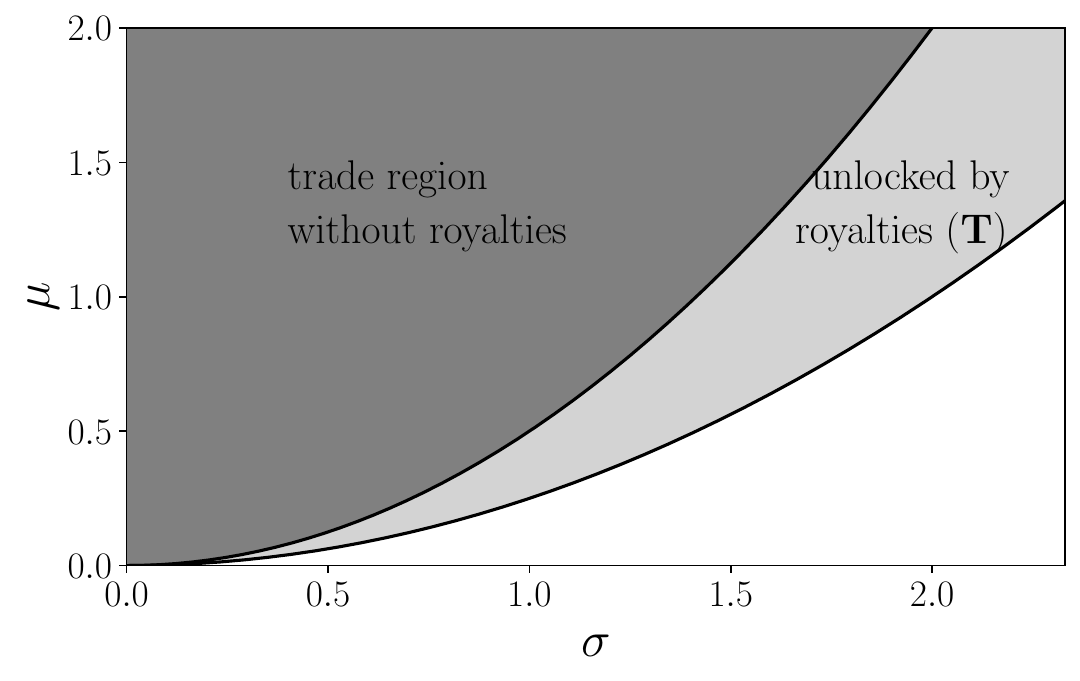}
    \caption{Regions of trade with and without royalties under risk aversion, in $\{\mu, \sigma\}$ space ($\eta_c=\eta_s=.5,c=0$).
    \label{fig:trade}}
\end{figure}

Figure~\ref{fig:trade} illustrates the Corollary~\ref{cor:TRrisk} trade region expansion in $\{\mu, \sigma\}$ space. The darker region depicts parameter combinations under which trade occurs regardless of whether royalties are available, while the lighter region~$\mathbbm{T}$ represents combinations where trade is feasible \emph{only} when royalties are enforced. The expansion occurs entirely in the high-variance portion of the parameter space: as $\sigma$ increases, the risk-sharing problem described above becomes more severe, and without royalties no price can simultaneously satisfy both parties' participation constraints. Royalties restore feasibility by allowing the creator to absorb a share of the resale variance, relaxing the speculator's participation constraint. As Figure~\ref{fig:trade} shows, the region~$\mathbbm{T}$ comprises a meaningful share of the total trade area, indicating that the market expansion enabled by royalties is economically significant.

\section{Royalties Under Information Asymmetry}
\label{sec:information_asymmetry}

In this section, we expand the base model in Section~\ref{sec:model} to allow information asymmetry. In practice, speculators may be better-informed about the prospects of the aftermarket \citep{ optimalpricingwithspeculators, defusco2022speculative}. To capture this, suppose the speculator has better information than the creator regarding customer willingness-to-pay. A simple way to model this advantage is to assume that the speculator can observe the realization of $V$ before deciding whether to purchase the NFT from the creator. This extreme form of informational advantage sharpens the contrast with the benchmark; a more realistic partial-information model would yield qualitatively similar but less analytically tractable results.\footnote{A more elaborate model would be to assume that speculator cannot perfectly observe buyer willingness to pay, but nonetheless retains an advantage over the creator, e.g.,  by receiving a more precise signal.} Creators, in turn, are aware of their disadvantage.

\subsection{Trade under Information Asymmetry Without Royalties}\label{sec:5.1}
Suppose for the moment that royalties are not allowed. In this case, the speculator's informational advantage collapses the outcome of the game for the creator, to that of Section~\ref{sec:benchdirect}, leaving her indifferent between selling to the speculator or to the end-buyer. To see this, consider that at time 1, if $p_0 > v$, the speculator refuses trade, and if $p_0 \leq v$, the speculator purchases the NFT, knowing that he can sell it for a profit $v-p_0 \geq 0$ to the end-buyer.

There are two implications. First, the speculator effectively extracts the full surplus from the end-buyer and realizes profit $v-p_0$. Second, the creator cannot extract all the revenue from either the speculator or the end-buyer.  In particular, the creator's expected revenues collapse to those absent the speculator described in Corollary~\ref{cor:nono}.

The creator's problem remains:
 \begin{equation*}
     \max_{p_0} p_0 \cdot \Pr [ V > p_0],
 \end{equation*}
and the previous result that the creator cannot extract maximum revenues, based on Markov's inequality,
\begin{equation*}
    p_0 \cdot \Pr [ V \ge p_0 ] \le \E[V], \quad \forall \ p_0 \in [0,\infty).
\end{equation*}
 
The following corollary recaps the main points.
\begin{corollary}[No Royalties, Better-Informed Speculator]\label{cor:infoab}
In the presence of a better-informed speculator, without the possibility of royalties, the creator's expected profits collapse back to those of Corollary~\ref{cor:nono}. The creator can no longer piggyback on the speculator's access to the aftermarket (described in Corollary~\ref{cor:noyes}).
\end{corollary}

\subsection{How Royalties Add Value under Information Asymmetry}

We now introduce royalties to the setup of Section~\ref{sec:5.1}, maintaining the speculator's informational advantage. Both the modeling modifications required, and the analysis, are more involved here.

As before, at time 0, the creator sets her mint price $p_0$ and royalty rate, $r$. At time 1, the speculator learns the valuation, $v$, and will purchase the NFT if and only if $(1-r)v - p_0 \geq 0$.
If the speculator does purchase the NFT, the creator's expected profits are given by $p_0 + r E\inbrak{V\suchthat (1-r)V \ge p_0}$, where the second term represents the royalties paid back to the creator once the speculator re-sells the NFT to an end-customer at time 2. Notice that the expectation is now conditional. This scenario occurs with probability $\Pr [ (1-r)V\geq p_0]$. On the other hand, if the speculator does not buy the NFT at time 1, the creator can still attempt to sell the NFT at time 2 to an end-buyer (at the mint price $p_0$). Trade in this latter scenario is conditional on two events occurring: if the end-buyer's valuation is high enough ($V \ge p_0$) and if the speculator did not purchase the NFT. The associated probability is $\Pr [ (1-r)\cdot V < p_0 \ \cap \ V \ge p_0]$.

Bringing these insights together, the creator's expected profits can be written as
\begin{equation}\label{eq:optinfoasymu}
     u_c = \Big ( p_0 + r E[V|(1-r)V \ge p_0] \Big ) \cdot \Pr [ (1-r)V\geq p_0] + p_0 \Pr [ (1-r)\cdot V < p_0 \ \cap \ V \ge p_0] - c.
\end{equation}

Rearranging terms (see details in the proof of Lemma~\ref{lem:optinfoasym}), the creator's maximization problem can be simplified to:
\begin{equation}
    \label{eq:optinfoasym}
    u_c^* = \max_{p_0, r} \quad p_0 (1 - F(p_0)) + r \int_{\frac{p_0}{1-r}}^\infty v f(v) dv - c.
\end{equation}
\begin{lemma}\label{lem:optinfoasym}
When a better-informed speculator is present, and royalties are allowed, the first order conditions associated with problem \eqref{eq:optinfoasym} are given by
\begin{align*}
    1-F(p_0) - p_0 f(p_0) &= p_0 \frac{1}{(1-r)^2} f\inparen{ \frac{p_0}{1-r} } \\
     \int_{\frac{p_0}{1-r}}^\infty v f(v) dv &= p_0^2\frac{ r}{(1-r)^3} f\inparen{\frac{p_0}{1-r}}.
\end{align*}
\end{lemma}
In general, solution uniqueness and closed-forms are not directly obvious in this case. Nonetheless, we have the following main result.

\begin{theorem}
\label{thm:asym}
    Under information asymmetry, royalties can fully restore the creator's ability to piggyback on the speculator's informational advantage. The creator can recover the maximum possible revenues $E[V]$ by giving out her NFT for free to the speculator ($p_0=0$), and relying exclusively on the future income generated by royalties. This yields her excess profits (compared to the case without royalties) of:
    $$\Delta = \E[V] - \max_{p_0} \ \Big ( p_0 \cdot \Pr[ V > p_0] \Big ) > 0.$$
\end{theorem}
\noindent The inequality is strict if $V$ is non-degenerate (Definition~\ref{def:non-degenerate}). This strategy of ``airdropping'' NFTs in the initial sale has sometimes been observed in practice. The general narrative  is that this generates buzz for the project. Theorem~\ref{thm:asym} gives an alternative, rational, explanation for this strategy.

Another interesting implication of Theorem~\ref{thm:asym} is that royalties can unlock value for creators by allowing them to bypass the inefficiencies associated with static pricing. It is as if creators could adapt their mint prices after uncertainty resolves to capture future price appreciation. In other words, royalties can be thought of as a roundabout way for creators to capture some of the benefits of \emph{adaptive pricing}. 

What's more, the magnitude of these benefits can be significant. To illustrate, we give below the benefit $\Delta$ for the exponential distribution. It will be useful to define the relative revenue increase  of including royalties (in \% of the suboptimal policy), which is given by
\begin{equation*}
    \mathcal{I} = \frac{\Delta}{ \max_{p_0} \ \Big ( p_0 \cdot \Pr[ V > p_0] \Big )}.
\end{equation*} 

\begin{figure}[ht]
        \centering
     \includegraphics[width=250pt]{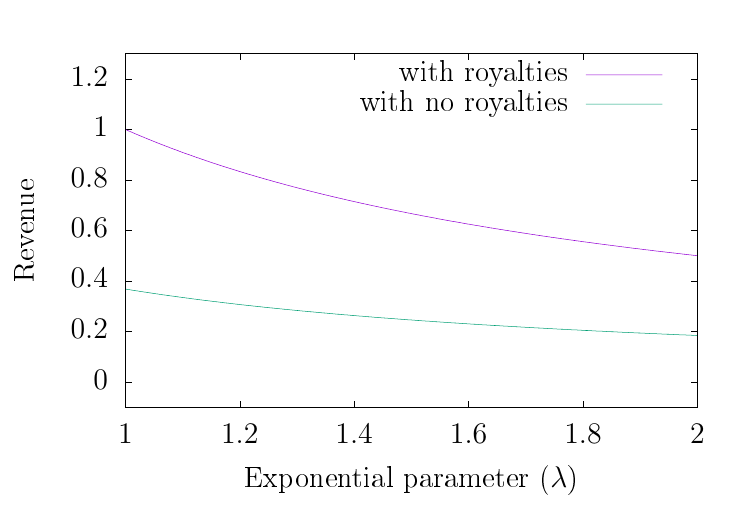}
     \caption{Creator revenue with optimal, $p_0,r$ compared to that without royalties (optimal $p_0$ when $r=0$).}
    \label{fig:asym_exponential_difference}
    \end{figure}

\begin{lemma}[Quantifying Gains, Exponential Distribution] \label{cor:asym_exponential}
Suppose $V\sim Exp[\lambda]$. Then solving the maximization problem with royalties in  \eqref{eq:optinfoasym} gives optimal creator revenues of $u^*_c = E[V]=\frac{1}{\lambda},$ whereas without royalties, the creator's optimal revenues are $u^*_{c,r=0}=\max_{p_0} \ \Big ( p_0 \cdot \Pr[ V > p_0] \Big ) = \frac{1}{e \lambda }$. This implies $\Delta = \frac{1}{\lambda}(1-\frac{1}{e})$, and $\mathcal{I}= e-1 \sim 1.72$, that is, the use of royalties increases creator revenues by $172\%, \ \forall  \lambda \in (1,\infty)$. This is illustrated in Figure~\ref{fig:asym_exponential_difference}.
\end{lemma}

{\bf Calibration to Data:}
    \cref{thm:asym} implies the optimal 
    royalty is $r = 1$. This extracts \emph{all} the revenue from the speculator.
    In practice, of course, we do not see royalty rates set to 100\%.
    Nevertheless, the point of this simplified model is to illustrate the inherent value that royalties could provide 
    to the creator, when the speculator has an informational advantage.

    If our objective were to calibrate the model to data, we could readily do so by considering two additional constraints that exist in practice:
    
    The first is simply that many NFT platforms  cap royalties, for instance at 10\%. Adding this constraint to the model would only slightly change the takeaway: the creator's new optimal strategy would be to set the royalty rate equal to the maximum possible amount allowed by the platform. 
    
    The second is that we have assumed speculators have zero outside option, and are willing to participate, even if in expectation they make zero. In practice, they would require a minimum amount of utility and/or expected return.
    In \cref{app:outsideoption}, we show that the if the speculator has an outside option (that can guarantee some minimal profit on their investment),
    then the creator can no longer set $r = 1$. Depending on the desired return, optimal royalty rates can be in the low single digits $\sim 5\%$, aligning with practice. The key is that even in that scenario, the creator still benefits from setting a positive royalty rate.

\subsubsection*{Royalties Expand Trade Under Information Asymmetry}
As was the case in Section~\ref{sec:riskaversion}, by increasing creator utility, royalties also expand the space of parameters in which trade occurs between creator and speculator. Let $\mathbbm{T}$ be the set of parameters for which trade would occur if royalties were allowed, but would not occur in their absence.
\begin{corollary}\label{cor:TRinfo}[Royalties expand the trade region under info. asymmetry]
Under information asymmetry, royalties increase the region of trade, that is, the set $\mathbbm{T} = \{\lambda, c \ | \ u_c^* \geq 0 \ \cap u_{c,r=0}^* < 0  \ \cap \lambda > 0 \ \cap  c > 0\}$ is non-empty. 
\end{corollary}

\begin{figure}[H]
    \centering
    \includegraphics[width=250pt]{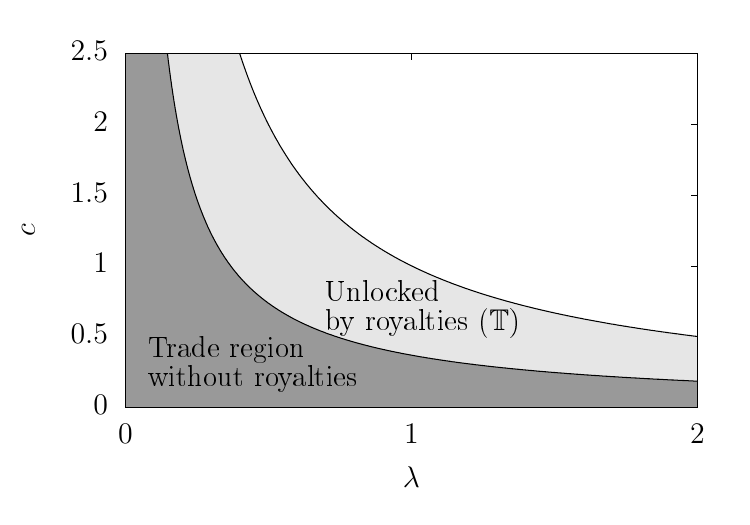}
    \caption{The region of trade unlocked by royalties under information asymmetry, in $\{\lambda, c \}$ space ($V\sim Exp[\lambda])$.
    \label{fig:tradeasym}}
\end{figure}

Figure~\ref{fig:tradeasym} illustrates the trade region expansion in
$\{\lambda, c\}$ space. The darker region depicts parameter combinations under
which trade occurs regardless of whether royalties are available, while the
lighter region~$\mathbbm{T}$ represents combinations where trade is feasible
\emph{only} when royalties are enforced. As minting costs increase and expected
end-buyer valuations fall (higher~$c$ and~$\lambda$, respectively), the surplus
available from the initial sale shrinks, eventually making trade infeasible
without royalties. Royalties restore feasibility by providing the creator with
a second source of revenue, contingent on the speculator's resale, that
offsets the thin margins from the primary transaction. As
Figure~\ref{fig:tradeasym} shows, this expansion comprises a substantial share
of the total trade area, indicating that the market-expanding role of royalties
is economically significant in this setting as well.

\section{Royalties as a Price Discrimination Mechanism}
\label{sec:price_discrimination}

In this section, we abstract away from risk aversion and information asymmetry, focusing instead on buyer heterogeneity in NFT collections. NFTs are sometimes sold as multi-unit collections to a pool of possible buyers, where each item of the collection can be identical or have some variations. 

\subsection{Modifications to base model} We expand our base model to capture a creator wishing to sell two NFTs to two end-buyers with uncertain (and independent) valuations. As before, a speculator is also present. The analysis in this discrete $2\times2$ case is already nontrivial: the game becomes combinatorial in nature. Nonetheless, this stylized case suffices to extract the main qualitative insights and shed some light on the economic forces at play.

At time 0, the creator sets a per-unit price, $p_0$, and royalty rate, $r$, for the collection, and at time 1, the speculator chooses whether to buy 0, 1 or 2 units at the price $p_0$. At time $2$, two end-buyers arrive with independent valuations $v_1,v_2 \sim V$.  The end buyers can buy from the creator (at price $p_0$), or from the speculator. As in \cite{optimalpricingwithspeculators}, we assume uncertainty is realized before the speculator acts in the secondary market. For simplicity,  the speculator can fully extract consumer surplus from end-buyers, meaning he can sell the NFT to buyer $i$ for the price $v_i$ (matching the buyer's willingness-to-pay). This represents a trading advantage which may be particularly salient in crypto markets as explained in \cite{daian2020flash}, where more sophisticated market players can arbitrage and front-run buyers and sellers in the block production process, even after trades have been submitted. 

Before proceeding, we emphasize that the assumption here differs from the information asymmetry assumption in Section~\ref{sec:information_asymmetry}. There, we assumed the speculator already knew the realization of buyer WTP before having to make the NFT purchase decision. Here, instead, the speculator makes the NFT purchasing decision without knowing the exact buyer WTP at time 0. His advantage is instead that he can  better extract WTP at time 2 from different buyers. So the two sections are capturing distinct effects.

One immediate observation in this setup is that if the speculator buys both units at time 1, he can sell them for prices $v_1,v_2$, so his expected revenue is $2E[V]$. Given there is no risk aversion and information asymmetry, one may wonder whether this multi-unit game collapses back to the base case of Theorem~\ref{thm:riskneutral}, where royalties can't add any extra value. More specifically, Theorem~\ref{thm:riskneutral} states that the creator could set the price $p_0=E[V]$ without using any royalties, and still extract the optimal level of revenues for herself. Similarly here, intuition may suggest that the creator could set $p_0=E[V]$ for each unit, and recover $2E[V]$ in total. However, this simple strategy fails because the speculator is not forced to buy all units in the collection!\footnote{The creator could try to force an all-or-nothing bundle upon the speculator, by embedding a clause in the smart contract. But this type of strategy isn't often observed in practice, likely because it would have non-trivial implications on end-buyer demand.} The speculator could, for instance, purchase only one of the two NFTs at time 1, and attempt to sell it to the end-buyer with the highest valuation at time 2. 

The latter option is in fact critical to the entire section. As we shall see, if price $p_0$ is in a certain range, the speculator indeed prefers to buy one unit rather than two. This possibility adds two sources of complexity into the analysis. First, one must account for the fact that the speculator will have to compete against the original seller at time 2, who in this scenario, would still have one unit on offer at the mint price $p_0$. In addition, to evaluate the potential benefits of price discrimination, we need to calculate the distribution of the maximum buyer valuation and the minimum buyer valuation. This makes the equilibrium of the multi-unit game non-trivial and motivates the definition of \emph{Order Statistics}.
\begin{definition}[Order statistics]
    Let $X_1,\ldots,X_n$ be random variables.
    Define $\ord{X}{1}{n},\ldots,\ord{X}{n}{n}$ to be the \emph{order statistics} of $X_1,\ldots,X_n$,
    in sorted order, i.e., $\ord{X}{1}{n} \le \ord{X}{2}{n} \le \cdots \le \ord{X}{n}{n}$.
\end{definition}

Though we only have two stochastic variables here, it is still useful to preserve this more general notation, given qualitative insights will hold for an arbitrary number of them.

With this definition $\ord{X}{1}{n} = \min(X_1,\ldots,X_n)$, and $\ord{X}{n}{n} = \max(X_1,\ldots,X_n)$.
Notice that $X_1 + \cdots + X_n = \ord{X}{1}{n} + \cdots + \ord{X}{n}{n}$ because the order statistics are just a reordering 
of the random variables.  In particular this implies that $2E[V] = E[\ord{V}{1}{2}] + E[\ord{V}{2}{2}]$, a property that will be useful in the analysis.

\subsection{Agent Revenues}
Next, we outline some key steps required in the analysis, that will be useful in (i) stating the creator's optimization problem over her profits and (ii) building intuition for the results. The detailed analysis and proofs are left for the appendix.

In order to solve the game, we proceed via backward induction, starting with the computation of the speculator's revenues and profits \emph{conditioned} on buying 0, 1 or 2 units, given by Lemma~\ref{lem:speculator-revenue}. We use the notation $\SR{n}$ to denote the speculator's revenue when they sell $n$ units to the end-buyers (ignoring, for now, the royalties that will subsequently need to be passed on to the creator),  and $\CR{n}$ to denote the creator's revenue when they sell $n$ units. Note, in the proof, we show that the number of units the speculator buys, depends on the price level $p_0$ set by the creator. This leads to three regions of interest which we can intuitively refer to as the ``low-price'' region (where the speculator buys both units), the ``mid-price'' region (where the speculator buys one unit), and the ``high-price'' region (where the speculator buys zero units). These regions are formally defined in the appendix.

\begin{lemma}[Speculator Revenue]
   \label{lem:speculator-revenue}
   The speculator's expected revenue and profit, conditioned on price regions, are as follows:

   \begin{center}
   \begin{tabular}{ccccc}
       \toprule
       \textbf{$p_0$ level} & \textbf{Units Bought} & \textbf{Rev. Label} & \textbf{Spec. Rev.($\SR{n}$)} & \textbf{Spec. Profit} \\
       \midrule
       High & 0 & $\SR{0}$ & $0$ & $0$ \\
       Medium & 1 & $\SR{1}$ & Eq. \eqref{eqn:speculator-revenue1} & $(1-r)\cdot\SR{1}-p_0$ \\
       Low & 2 & $\SR{2}$ & $2 \cdot E[V]$ & $2\cdot\left((1-r)\cdot E[V]-p_0\right)$ \\
       \bottomrule
   \end{tabular}
   \end{center}

   where revenue $\SR{1}$ is given by:
   \begin{align}
       \SR{1} = \;E \inbrak{ \ord{V}{2}{2} \suchthat \ord{V}{2}{2} < p_0 } \cdot \left(F_V(p_0)\right)^2  &+ 2 \cdot p_0 \cdot F_V(p_0) \cdot \left( 1 - F_V(p_0) \right) \nonumber\\
        &+ E \inbrak{ \ord{V}{2}{2} \suchthat \ord{V}{2}{1} > p_0 } \cdot \left( 1 - F_V(p_0) \right)^2. \label{eqn:speculator-revenue1}
   \end{align}
\end{lemma}
Next, Lemma~\ref{lem:creator-revenue-multi} gives the analogous result for the creator's revenues (the creator's profits are omitted as they only differ by a constant, cost $c$).
\begin{lemma}[Creator Revenue]
    \label{lem:creator-revenue-multi}
    The creator's expected revenue is summarized as follows:

    \begin{center}
    \begin{tabular}{ccc}
        \toprule
        \textbf{$p_0$ Level} & \textbf{Rev. Label} & \textbf{Creator Revs. ($\CR{k}$)} \\
        \midrule
        High & $\CR{0}$ & $2 \cdot p_0 \cdot (1 - F_V(p_0))$ \\
        Medium & $\CR{1}$ & $p_0 + r \cdot \SR{1} + p_0 \cdot \Pr[\ord{V}{1}{2} > p_0]$ \\
        Low & $\CR{2}$ & $2 \cdot (p_0 + r \cdot E[V])$ \\
        \bottomrule
    \end{tabular}
    \end{center}
\end{lemma}
Next, we look into the equilibrium of the game with and without royalties.

\subsection{Trading Multi-Unit NFTs Without Royalties}

The creator's optimization problem is to maximize her expected revenues described in Lemma~\ref{lem:creator-revenue-multi} over $p_0$ and $r$. 
Denote these $u_c^*$ (at the optimal $p_0^*,r^*$) and let $u_{c,r=0}^*$ represent the creator's optimal revenues without royalties, that is, the optimal $p_0^*$ when $r=0$. 

To make further progress from here, it will be helpful to assume a distribution for $V$, as some of the conditions and results are distribution-specific. This will also help us illustrate the results visually. Here and below, we thus assume $V$ is exponentially distributed (the qualitative results continue to hold with other standard distributions as well).

Before fully solving the problem, we can already state the following result, showing some limits to the maximal achievable creator revenues, when not using royalties.
\begin{theorem}
    \label{thm:multi-unitr0}
    If $V$ is exponentially distributed with parameter $\lambda$, and royalties are ruled out, the creator selling two units \emph{cannot} extract the maximum possible revenue, 
    that is,  $u^*_{c,r=0} < 2E[V]$.
\end{theorem}
In the next section, we characterize the optimal solution and objective function when royalties are allowed, as well as the gap $\Delta$ that they generate compared to the case without royalties.

\subsection{How Royalties Add Value in Multi-Unit Settings}

In this section, we show that when creators are selling multi-unit collections, they can leverage royalties to piggyback on speculators' ability to price discriminate end-buyers. The creator can recover the maximum possible revenues by airdropping her NFTs for free to the speculator ($p_0=0$), and relying exclusively on the future income generated by royalties. This strategy extracts the highest possible revenues to the creator, $u^*_c = 2 E[V]$ and yields her excess profits (compared to the case without royalties) of:
$$\Delta = 2\E[V] - u_{c,r=0}^* > 0.$$

\noindent We emphasize that the result is driven by neither risk aversion nor information asymmetry.\footnote{Unlike Section~\ref{sec:information_asymmetry} where the speculator observes the valuation ex ante before the purchase decision, here, the valuation is learned only ex post.} Rather, it is a mere consequence of the fact that royalties allow the creator to piggyback on the speculator's access to the aftermarket, and his ability to charge different buyers different prices. In other words, royalties can be thought of as a roundabout way for creators to capture the benefits of {\bf price discrimination}, even as collection units {\bf are all sold at the mint price}.

As before, the magnitude of these benefits can be meaningful. To illustrate, we give below the benefit $\Delta$ for the exponential distribution. The relative revenue increase  of including royalties (in \% of the suboptimal policy), is given by
\begin{equation*}
    \mathcal{I} = \frac{\Delta}{ u^*_{c,r=0} }.
\end{equation*} 

\begin{lemma}[Quantifying Gains, Exponential Distribution] \label{lem:multi_exponential}
Suppose $V\sim Exp[\lambda]$. Then solving the maximization problem with royalties in  \eqref{eq:optinfoasym} gives optimal creator revenues of $u^*_c = E[V]=\frac{1}{\lambda},$ whereas without royalties, the creator's optimal revenues are $u^*_{c,r=0} < .971 \cdot 2 \cdot E[V]$. This implies $\Delta \ge .029 \cdot 2 \cdot E[V]$, and $\mathcal{I} \sim .029$, that is, the use of royalties increases creator revenues by $2.9\%, \ \forall  \lambda \in (1,\infty)$. 
\end{lemma}

\subsubsection*{Royalties Expand Trade for
NFT Collections}
\begin{figure}[H]
    \centering
    \includegraphics[width=250pt]{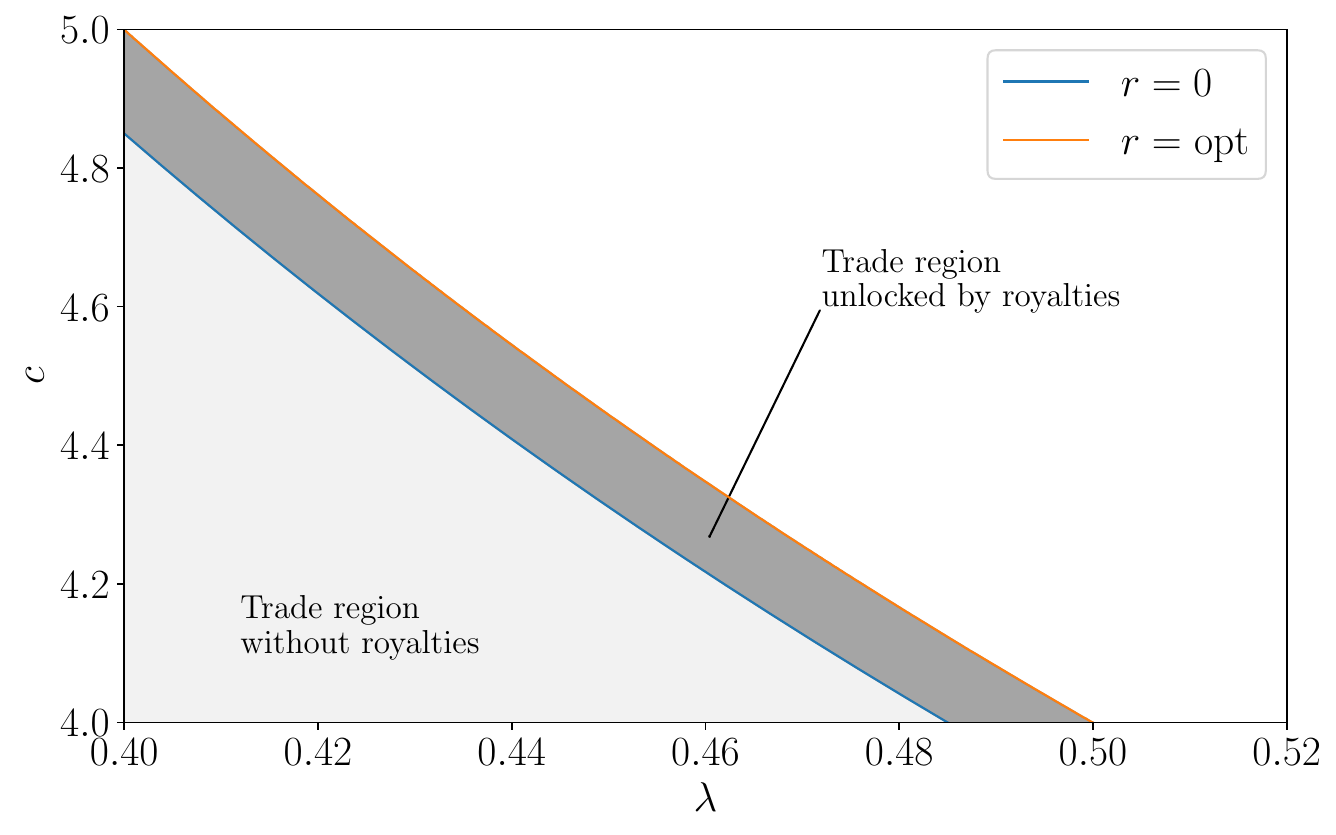}
    \caption{The region of trade unlocked by royalties when selling NFT Collections, in $\{\lambda, c \}$ space ($V\sim Exp[\lambda])$.
    \label{fig:trademulti}}
\end{figure}

As was the case in Sections~\ref{sec:riskaversion} and \ref{sec:information_asymmetry}, by increasing creator utility, royalties also expand the space of parameters in which trade occurs between creator and speculator. Let $\mathbbm{T}$ be the set of parameters for which trade would occur if royalties were allowed, but would not occur in their absence.
\begin{corollary}\label{cor:TRcollections}[Royalties expand the trade region for NFT Collections]
When selling collections of multiple NFTs, royalties increase the region of trade, that is, the set $\mathbbm{T} = \{\lambda, c \ | \ u_c^* \geq 0 \ \cap u_{c,r=0}^* < 0  \ \cap \ \lambda > 0 \ \cap  c > 0\}$ is non-empty. A visual representation is provided in Figure~\ref{fig:trademulti}.
\end{corollary}

\section{Discussion}
\label{sec:discussion}

Creator royalties generate meaningful economic value, not despite the presence of sophisticated speculators, but because of them. Embedding an optimized royalty clause in a friction-rich model reveals that royalties gain power from the very intermediaries often blamed for neutralizing them.

\emph{Contributions to the literature:} We make three contributions. First, we establish a formal irrelevance benchmark for creator royalties---analogous to Modigliani--Miller for capital structure---showing that royalties have no effect in frictionless markets with equally-informed, risk-neutral agents. Just as Modigliani--Miller irrelevance motivates the study of which frictions make capital structure matter, this benchmark identifies precisely which market imperfections give royalties their bite. Second, we show that each of three common imperfections---differential risk tolerance, information asymmetry, and buyer heterogeneity---independently transforms royalties from neutral transfers into value-creating instruments that allow creators to work around these imperfections, and that in all three cases royalties expand the region of feasible trade. These benefits persist even when rational buyers fully anticipate royalty costs. Third, we provide a theoretical account of the empirical patterns documented by \citet{Tunc2024NFTRoyalty}: our overconfidence extension (Appendix~\ref{app:overconfidence}) generates the negative correlation between royalty rates and creator revenue they observe, while our participation-margin results explain why the option to set royalties remains valuable despite this correlation.

\emph{Practical Implications:} Creators benefit from royalties not only as passive income streams but as strategic instruments that harness speculative market dynamics. The emergence of enforcement technologies like ERC-721C, MIP-1, and protocol-level implementations makes this increasingly relevant: creators now have credible tools to ensure royalty collection.

These results also carry implications for platforms. Because royalties enlarge the set of viable transactions (Corollaries~\ref{cor:TRrisk}, \ref{cor:TRinfo}, and~\ref{cor:TRcollections}), they can attract creators who would not otherwise participate---expanding supply and potentially improving content quality and user engagement \citep{DeMatteo2023}. Platforms that offer effective royalty enforcement may therefore gain a competitive advantage in attracting committed creators with established collector bases. At the same time, royalties act as a tax on future transfers, which can dampen secondary-market activity. As formalized in Appendix~\ref{app:liquidity}, sufficiently adverse liquidity effects can make the optimal royalty rate zero, highlighting the tension platforms face between creator retention and buyer-side incentives. A full treatment of platform incentives and inter-platform competition falls outside our scope but represents a valuable direction for future research. More broadly, policymakers should regard royalties as efficiency-enhancing mechanisms that promote sustainable creator economies, not merely redistributive transfers.

\emph{Limitations and Future Directions:} Our model makes several deliberate simplifying choices to isolate the economic forces at play. Extending the framework to incorporate multiple competing speculators/buyers is a natural direction, partially addressed in our multi-unit analysis in Section~\ref{sec:price_discrimination}, but a full competitive model could yield additional insights into how speculator competition affects the creator's optimal royalty rate. Second, we model speculators as purely profit-motivated intermediaries, consistent with the speculation literature \citep{optimalpricingwithspeculators, Courty2003}, and do not endow them with hedonic value for the NFTs themselves. Third, our model uses a ``royalties not allowed'' benchmark, which is distinct from a setting where royalties exist but are not enforced; the latter would permit voluntary compliance (akin to tipping) and represents a different economic environment. Fourth, our analysis focuses on individual collections and does not account for spillover effects across collections or platforms. Finally, other topics such as search costs, network effects, and dynamic creator learning remain unexplored. Each of these represents a promising avenue for future research.

Despite these simplifications, the core insight is robust. Far from being a relic of legacy copyright law, royalties in frictional digital markets are sophisticated, programmable instruments that let creators harvest the very speculative energy that critics claim will undermine them. As enforcement moves from marketplace-dependent suggestion to creator-controlled mechanism, royalties can realign incentives between artists, traders, and platforms---reviving the centuries-old ideal that creators should share in the upside of their own success.
\begin{spacing}{1.45}
\bibliography{main}
\end{spacing}

\appendix

\newpage
\begin{center}
    \LARGE{Appendix (attached to paper)}
\end{center}

\section{Auxiliary results}

\subsection{Are Royalties Actually Paid? Empirical Evidence}\label{sec:empirical}

We examine royalty enforcement using transaction data from three major marketplaces: LooksRare, OpenSea, and X2Y2. Data were obtained via Alchemy's \href{https://docs.alchemy.com/reference/getnftsales}{getNFTSales} API, providing comprehensive transaction records for each platform.

\textbf{LooksRare.} Covers December 29, 2021 to April 13, 2023, with 401,636 sales (186 MB). Of these, 260,547 transactions ($\approx 64.88\%$) include royalty payments.

\textbf{OpenSea.} Combines Seaport and Wyvern protocols.\footnote{There is no duplication between Wyvern and Seaport records, as OpenSea transitioned from Wyvern to Seaport during mid-2022; at any point in time, only one protocol was active.} Seaport spans June 12, 2022 to March 20, 2024, with 13,686,195 sales and 11,485,580 ($\approx 83.94\%$) including royalties (6.4 GB). Wyvern covers February 18, 2022 to August 1, 2022, with 7,911,472 sales and 7,481,619 ($\approx 94.56\%$) royalty-enforcing transactions (3.8 GB).

\textbf{X2Y2.} From February 4, 2022 to March 21, 2024, includes 1,897,105 sales (868 MB), all with enforced royalties (100\%).

Across all platforms, 99.64\% of transactions used ETH or wrapped ETH; we exclude others for consistency. Table~\ref{tab:royalty_data} summarizes key statistics.

\begin{table}[ht]
\centering
\begin{tabular}{@{}lccc@{}}
\toprule
\textbf{Metric} & \textbf{OpenSea} & \textbf{LooksRare} & \textbf{X2Y2} \\
\midrule
Avg Royalty Rate & 7.90\% & 6.40\% & 0.67\% \\
Median Royalty Rate & 8.33\% & 5.38\% & 0.00\% \\
Std Dev of Rates & 3.11\% & 3.92\% & 2.03\% \\
Avg ETH Royalty Paid & 9.85 ETH & 5.09 ETH & 0.47 ETH \\
Total ETH Royalties & 332,593.44 & 21,236.38 & 5,892.08 \\
\bottomrule
\end{tabular}
\caption{Royalty Statistics Across Marketplaces}
\label{tab:royalty_data}
\end{table}

Royalty enforcement relies on platforms. While creators may specify royalties in contracts (e.g., via ERC2981), these contracts lack contextual awareness (e.g., sale price) and cannot autonomously enforce payments.

\begin{figure}[ht]
    \centering
    \begin{subfigure}{0.32\textwidth}
        \includegraphics[width=\linewidth]{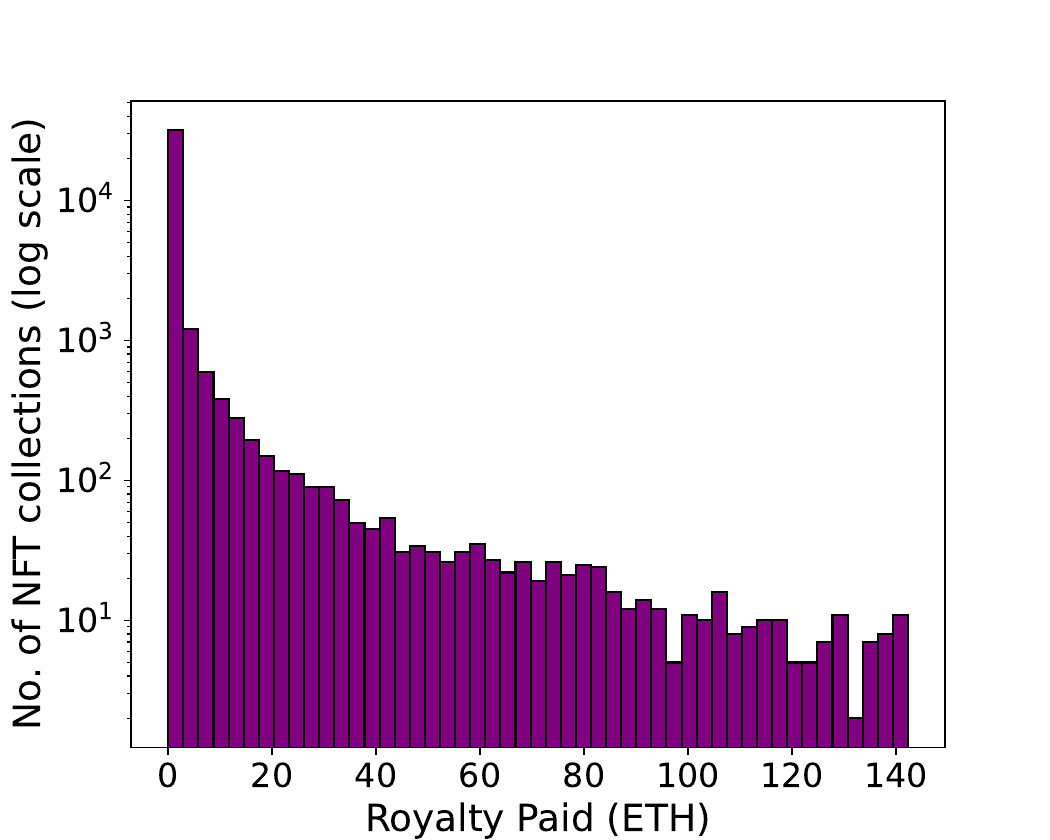}
        \caption{Royalties paid per collection}
        \label{fig:sub_royaltypay}
    \end{subfigure}
    \hfill
    \begin{subfigure}{0.32\textwidth}
        \includegraphics[width=\linewidth]{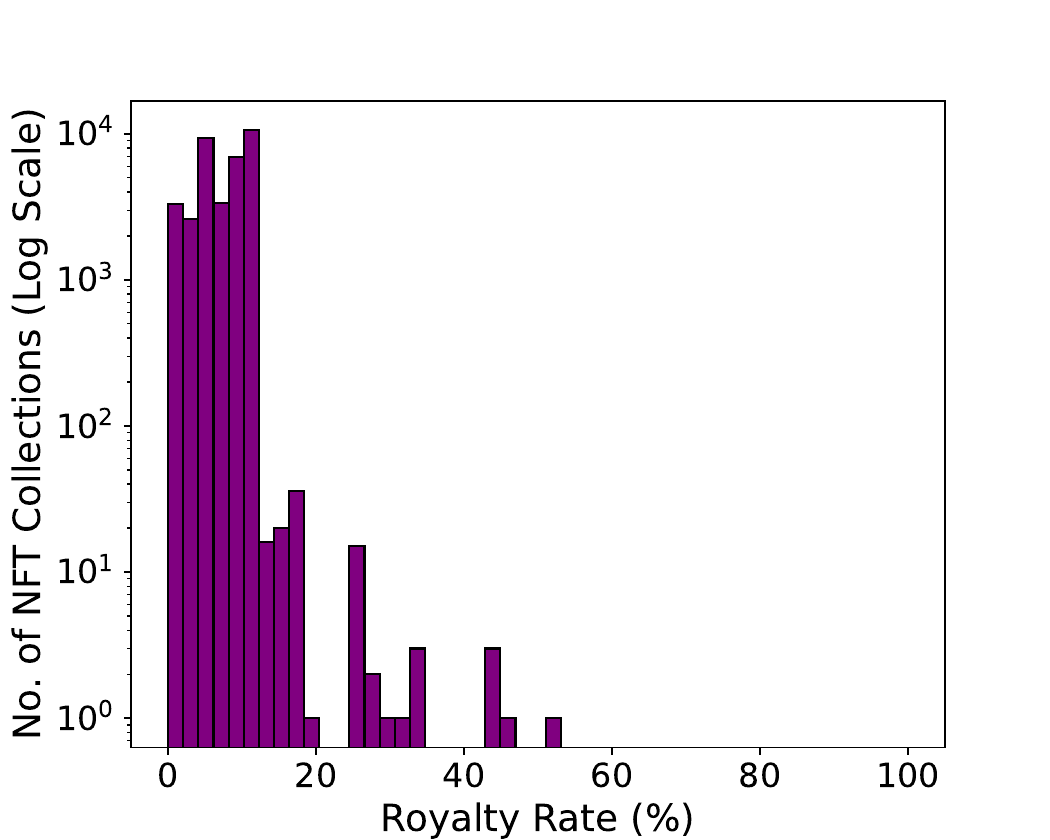}
        \caption{Royalty rate distribution}
        \label{fig:sub_rate_dist}
    \end{subfigure}
    \hfill
    \begin{subfigure}{0.32\textwidth}
        \includegraphics[width=\linewidth]{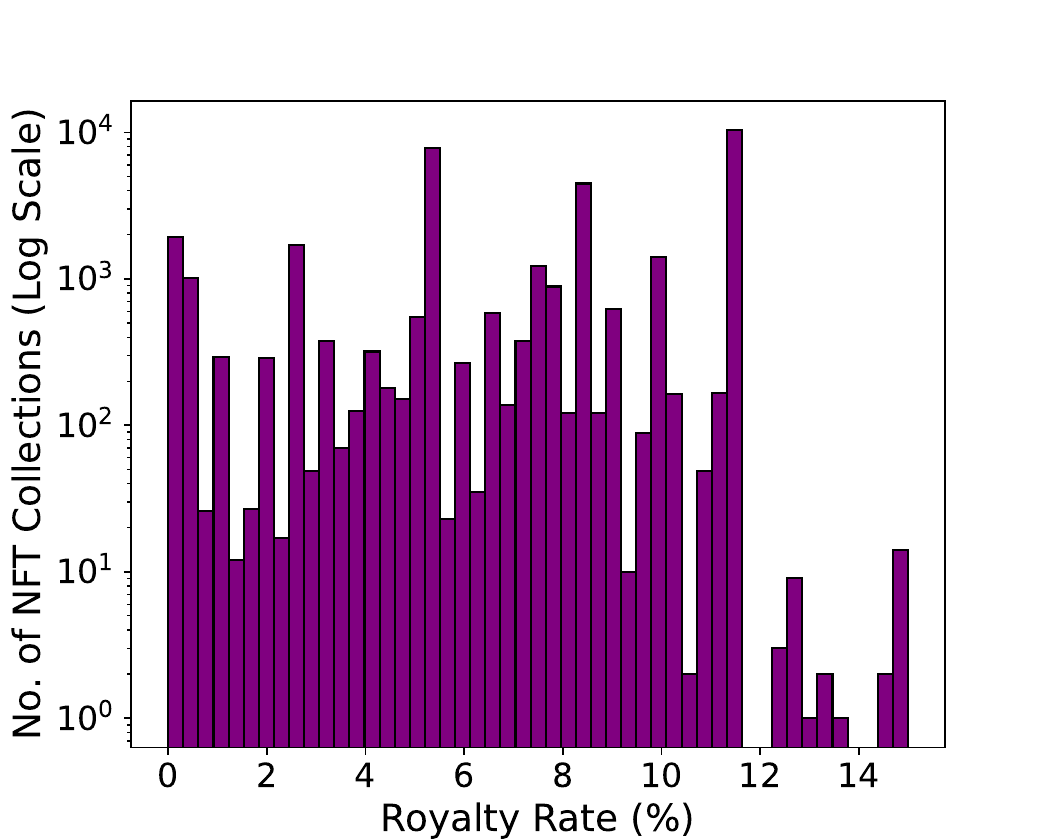}
        \caption{Rate Zoomed: 0–15\% Range}
        \label{fig:sub_rate_zoom}
    \end{subfigure}
    \caption{Aggregated NFT Royalty Enforcement Across Marketplaces}
    \label{fig:combined_royalty_figures}
\end{figure}

Figure~\ref{fig:combined_royalty_figures} presents aggregate royalty statistics. Panel (a) shows the ETH royalty paid per collection (n=36,224). 
 The mean is 9.93 ETH (SD = 183.66), with a median of 0.036 ETH; 75\% of collections receive less than 0.44 ETH. The maximum exceeds 21,000 ETH ($\sim$76M USD).

Panels (b) and (c) display enforced royalty rate distributions. The mean rate is 7.35\% (SD = 3.63\%), with quartiles [5.41\%, 11.43\%]. Approximately 4.7\% of collections have a 0\% rate. Most fall under 15\%, justifying the zoom in panel (c).


\subsection{Non-Degenerate Distribution}\label{app:6pages-nondegen}

\begin{definition}
	\label{def:non-degenerate}
	A random variable, $V$, is called \emph{non-degenerate} if $V$ is supported on at least two nonzero points, i.e., 
	there exists points $0 < v_1 < v_2$ and a value $\eps$, with $0 < \eps < (v_2-v_1)/2$, such that 
	$\Pr [ v_1 - \eps < V < v_1 + \eps ] > 0$, and $\Pr[ v_2 - \eps < V < v_2 + \eps ] > 0$.
\end{definition}

\begin{lemma}
	\label{lem:markovloose}
	If $V$ is a non-negative, non-degenerate random variable (\cref{def:non-degenerate}), then Markov's inequality is strict, i.e.,
	\begin{equation*}
		v \Pr [V>v] < E[V]
	\end{equation*}
	for all $v$.
\end{lemma}

\begin{proof}[\bf Proof of Lemma~\ref{lem:markovloose}]
	Since $V$ is non-degenerate, there exists points $0 < v_1 < v_2$ and a value $\eps$, with $0 < \eps < (v_2-v_1)/2$, such that 
	$\Pr [ v_1 - \eps < V < v_1 + \eps ] > 0$, and $\Pr[ v_2 - \eps < V < v_2 + \eps ] > 0$

	Let $\delta \defined \min\inparen{ \Pr [ v_1 - \eps < V < v_1 + \eps ], \Pr[ v_2 - \eps < V < v_2 + \eps ] }$.  Then, $\delta > 0$.

	Now, consider two cases:
	\textbf{Case 1:} If $v \le v_1+\eps$, then 
	\begin{align*}
		E[V] - v \cdot \Pr[V>v] &= \int_0^\infty x f_V(x) dx - v \int_{v}^\infty f_V(x) dx \\
								&\ge \int_{v_2 -\eps}^{v_2+\eps} (x-v_1) f_V(x) dx \\
								&\ge (v_2-\eps-v_1) \int_{v_2 -\eps}^{v_2+\eps} f_V(x) dx \\
								&\ge (v_2-\eps-v_1) \delta \\
								&> 0
	\end{align*}
	\textbf{Case 2:} If $v_1 + \eps < v$, then 
	\begin{align*}
		E[V] - v \cdot \Pr[V>v] &= \int_0^\infty x f_V(x) dx - v \int_{v}^\infty f_V(x) dx \\
								&\ge \int_0^{v_1+\eps} x f_V(x) dx \\
								&\ge \int_{v_1-\eps}^{v_1+\eps} x f_V(x) dx \\
								&\ge (v_1-\eps) \delta \\
								&> 0.
	\end{align*}
\end{proof}

\subsection{Robustness under Non-Zero Outside Options}\label{app:outsideoption}

In \cref{sec:information_asymmetry}, we found that the creator's optimal strategy was to set a price $p_0 = 0$, and a royalty rate $r = 1$.  In this setting, the speculator's profit was zero.

If we give the speculator an outside option, then the speculator will only participate if they are guaranteed some minimal amount of profit, and the creator can no longer set the royalty rate to 100\%.

In this section, we imagine the speculator has an outside option, and will only participate if their profit is at least
\begin{equation}
    \label{eq:outside_option}
    \Oopt(p_0) \defined  a_0 + a_1 \cdot p_0,
\end{equation}
which represents a minimum fixed external utility ($a_0$) and a minimum return required on the purchase price ($a_1$). With this constraint, the creator's revenue is given by

\begin{equation}
    \label{eq:creator_profit_with_outside_option}
    u_c^* \defined  \max_{p_0,r} \inparen{p_0 \cdot (1 - F(p_0)) + r \cdot \int_{\frac{\Oopt(p_0)}{1-r}}^{\infty} v \cdot f(v) \, dv}
\end{equation}

Compared to \cref{eq:optinfoasym}, the only change is that the lower bound of the integral is now $\frac{\Oopt(p_0)}{1-r}$ instead of $\frac{p_0}{1-r}$. Alternatively, we can express the creator's revenue as

\begin{equation}
    \label{eq:creator_profit_with_outside_option_alt}
    p_0 \cdot \Pr \inbrak{ V \ge p_0 } + r \cdot E \inbrak{ V \suchthat V \ge \frac{\Oopt(p_0)}{1-r} } \cdot \Pr \inbrak{ V \ge \frac{\Oopt(p_0)}{1-r} }
\end{equation}

Compare \cref{eq:creator_profit_with_outside_option_alt} to \cref{eqn:altuinfoasym}. As an example, suppose $V$ is uniform on $[a,b]$.  Then
\begin{align*}
    \Pr \inbrak{ V \ge p_0 } &= \frac{b-p_0}{b-a} \quad \mbox{ (for $a \le p_0 \le b$)} \\
    \Pr \inbrak{ V \ge \frac{\Oopt(p_0)}{1-r} } &= \frac{b-\frac{h(p_0)}{1-r}}{b-a} \quad \mbox{ (for $a \cdot(1-r) \le h(p_0) \le b \cdot (1-r)$)} \\
    E \inbrak{ V \suchthat V \ge \frac{\Oopt(p_0)}{1-r} } &= \frac{1}{2} \cdot \inparen{ \frac{h(p_0)}{1-r} + b } \quad \mbox{ (for $a \cdot(1-r) \le h(p_0) \le b \cdot (1-r)$)}
\end{align*}

So the creator's revenue is given by 
\begin{align}
    &p_0 \cdot \inparen{ \frac{ b-p_0}{b-a}} +  \frac{1}{2} r \cdot \inparen{ \frac{h(p_0)}{1-r} + b} \inparen{ \frac{b- \frac{h(p_0)}{1-r}}{b-a}} \\
    &= \frac{1}{b-a} \inparen{ p_0 \cdot (b-p_0) +  \frac{1}{2} r \cdot \inparen{ b^2 - \inparen{\frac{h(p_0)}{1-r}}^2}}
    \label{eqn:optimal_outside}
\end{align}

\begin{figure}[ht]
    \centering
    \includegraphics[width=0.8\textwidth]{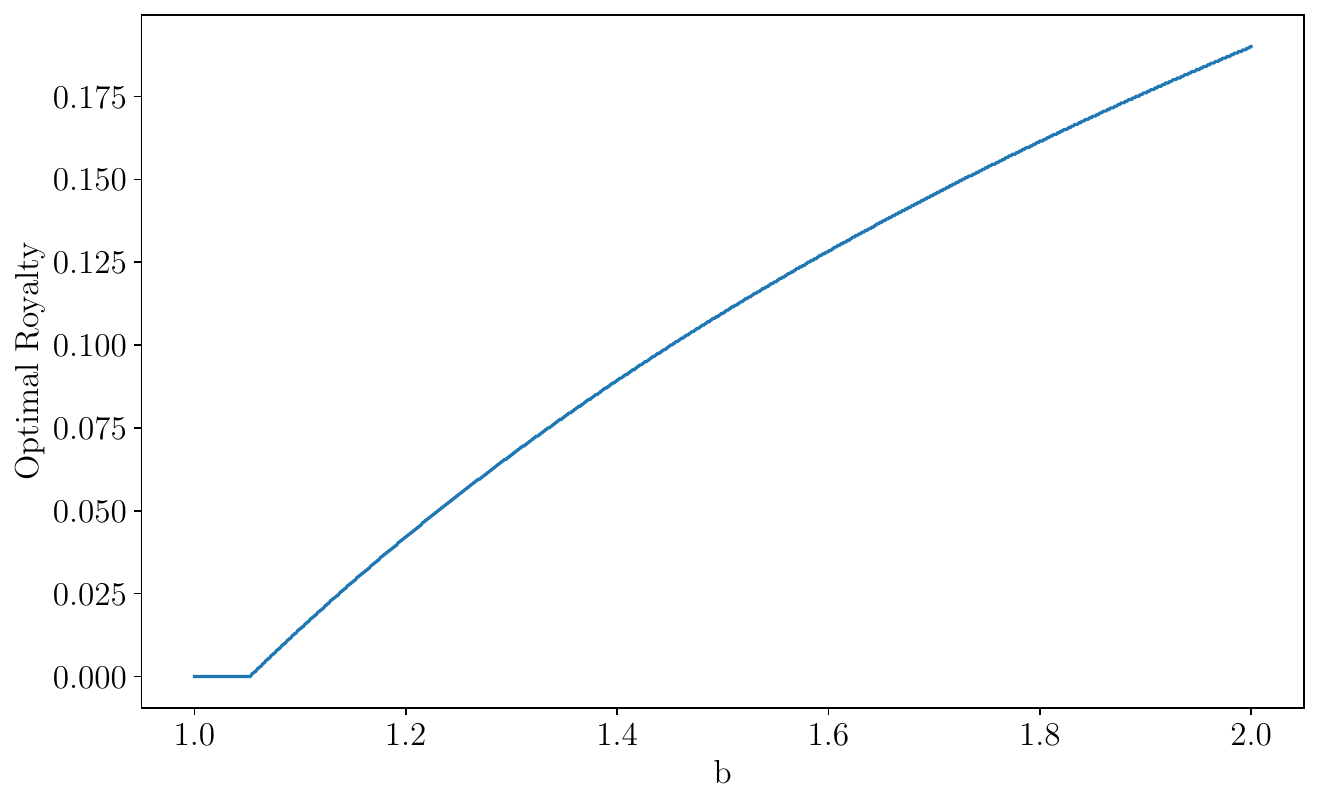}
    \caption{The optimal royalty rate as a function of $b$ when $V$ is uniformly distributed on $[0,b]$.
    Note that when the speculator has an outside option, the optimal royalty rate is strictly less than $100\%$.
    In this plot, the speculator's outside option is $\Oopt(p_0) \defined .5 + 1.05 \cdot p_0$.
    Note the flat part of the curve (for low values of $b$) occurs because if $b$ is low enough, the speculator's outside option means that the speculator is unwilling to purchase the item at \emph{any} royalty rate.  This plot is obtained by numerically optimizing \cref{eqn:optimal_outside}.}
    \label{fig:oo_optimal_uniform_royalty}
\end{figure}

Maximizing this expression over $p_0$ and $r$ is involved and does not yield simple closed-form solutions. Nonetheless,  we can solve the problem numerically. Setting $a = 0$, \cref{fig:oo_optimal_uniform_royalty} shows the optimal royalty rate as a function of $b$.
The key takeaway is that when the speculator has an outside option, the creator's optimal royalty rate is strictly less than $100\%$.

\begin{figure}[ht]
    \centering
    \includegraphics[width=0.8\textwidth]{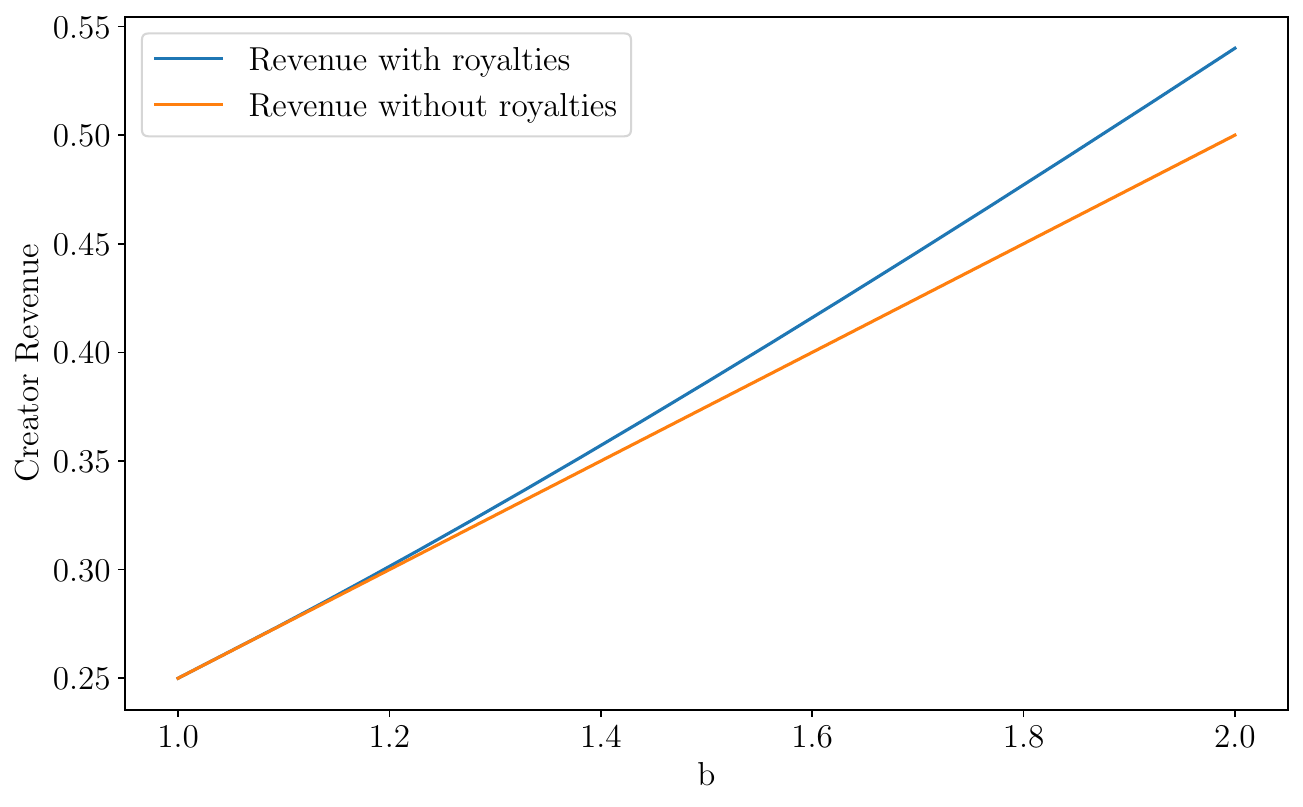}
    \caption{Even when the speculator has an outside option, the creator still benefits from setting a positive royalty rate.
    In this plot, as in \cref{fig:oo_optimal_uniform_royalty}, the buyer's valuation, $V$, is uniform on $[0,b]$, and the speculator's outside option is $\Oopt(p_0) \defined .5 + 1.05 \cdot p_0$.  The ``revenue with royalties'' line shows the creator revenue with the optimal royalty (as calculated in \cref{fig:oo_optimal_uniform_royalty}).
    The two curves overlap for small values of $b$ because at these values the speculator is unwilling to purchase the NFT (even with 0 royalties), and thus the creator's revenue is the same whether or not they have the option to charge royalties.  As in \cref{fig:oo_optimal_uniform_royalty} the optimal royalty is calculated by numerically optimizing over all possible royalty rates.}
    \label{fig:oo_revenue_with_and_without_royalties}
\end{figure}

\cref{fig:oo_revenue_with_and_without_royalties} shows the creator's revenue as a function of $b$.
The key takeaway is that even when the speculator has an outside option, the creator still benefits from setting a positive royalty rate.

\subsection{Robustness under Overconfidence}\label{app:overconfidence}

Our starting point is the setting in Appendix~\ref{app:outsideoption}.  Following \cite{Tunc2024NFTRoyalty}, we adopt a standard reduced-form approach to optimism/overconfidence as a distortion in beliefs about payoff-relevant fundamentals: the creator chooses $(p_0,r)$ to maximize \emph{perceived} expected revenue under a subjective distribution that first-order stochastically dominates the true one, but realized revenue is evaluated under the true distribution (e.g., \citealp{Heaton2002Optimism,LandierThesmar2009Optimistic}). 

\paragraph{Belief distortion.}
Let $V\sim F$ with density $f$.
An overconfident creator with parameter $\theta\ge 1$ perceives valuations as inflated by factor $\theta$, i.e.,
\begin{equation}\label{eq:oc_distortion_full}
\widehat F_\theta(v)\defined F\!\inparen{\frac{v}{\theta}},
\qquad
\widehat f_\theta(v)\defined \frac{1}{\theta} f\!\inparen{\frac{v}{\theta}},
\end{equation}
where $\widehat f_\theta(v)=\frac{d}{dv}\widehat F_\theta(v) = \frac{d}{dv} F(v/\theta)
= f(v/\theta)\cdot \frac{d}{dv}\inparen{\frac{v}{\theta}} = \frac{1}{\theta}f(v/\theta)$ by the chain rule. 

Equivalently, $\widehat F_\theta$ is the law of $\widehat V \defined \theta V$, since
$\Pr(\widehat V \le v)=\Pr(V \le v/\theta)=F(v/\theta)$. Thus, belief distortion corresponds to a multiplicative scale inflation of valuations.

\paragraph{Choice under biased beliefs and realized revenue.}
Given $(p_0,r)$, perceived creator revenue is
\begin{equation}\label{eq:oc_perceived_objective_full}
\widehat u_c(p_0,r;\theta)
\defined
p_0 \cdot \inparen{1-\widehat F_\theta(p_0)}
+
r \cdot \int_{\frac{\Oopt(p_0)}{1-r}}^{\infty} v \cdot \widehat f_\theta(v)\, dv,
\end{equation}
while realized revenue is
\begin{equation}\label{eq:oc_true_objective_full}
u_c(p_0,r)
\defined
p_0 \cdot \inparen{1-F(p_0)}
+
r \cdot \int_{\frac{\Oopt(p_0)}{1-r}}^{\infty} v \cdot f(v)\, dv.
\end{equation}
Let $(p_0(\theta),r(\theta))\in \arg\max_{p_0,r} \widehat u_c(p_0,r;\theta)$ denote the creator’s choice under overconfidence, and let $(p_0(1),r(1))$ be the unbiased benchmark.

\paragraph{Uniform illustration with outside option.}
To keep the extension maximally transparent and aligned with \cref{app:outsideoption}, consider the same uniform specification and fix $p_0=0$.
Let $V\sim \mathrm{Unif}[0,b]$ and $\Oopt(0)=a_0$, with $0<a_0<b$.
For any $r\in[0,1)$ such that $\frac{a_0}{1-r}\le b$, true revenue equals
\begin{equation}\label{eq:oc_uniform_true_revenue_full}
u(r;b)
\defined
r\cdot \int_{\frac{a_0}{1-r}}^{b} v \cdot \frac{1}{b}\, dv
=
\frac{r}{2b}\inparen{b^2 - \inparen{\frac{a_0}{1-r}}^2}.
\end{equation}

An overconfident creator with parameter $\theta\ge 1$ behaves as if valuations were $\mathrm{Unif}[0,\theta b]$. To see this, consider
\begin{align}
    V \sim \mathrm{Unif}[0,b],\quad F(v) = \begin{cases}
        v\cdot \frac{1}{b}, &v\in[0,b] \\
        0, &v \not\in[0,b]
    \end{cases} \implies \hat{F}_\theta(v) = F\inparen{\frac{v}{\theta}} = \begin{cases}
        v\cdot \frac{1}{\theta b}, &v\in[0, \theta b] \\
        0, &v\not\in[0,\theta b]
    \end{cases}
\end{align}

Thus, the perceived revenue is
\begin{equation}\label{eq:oc_uniform_perceived_revenue_full}
\widehat u(r;b,\theta)
\defined
r\cdot \int_{\frac{a_0}{1-r}}^{\theta b} v \cdot \frac{1}{\theta b}\, dv = 
\frac{r}{2\theta b}\inparen{(\theta b)^2 - \inparen{\frac{a_0}{1-r}}^2}.
\end{equation}

\begin{figure}[H]
    \centering
    \includegraphics[width=0.8\textwidth]{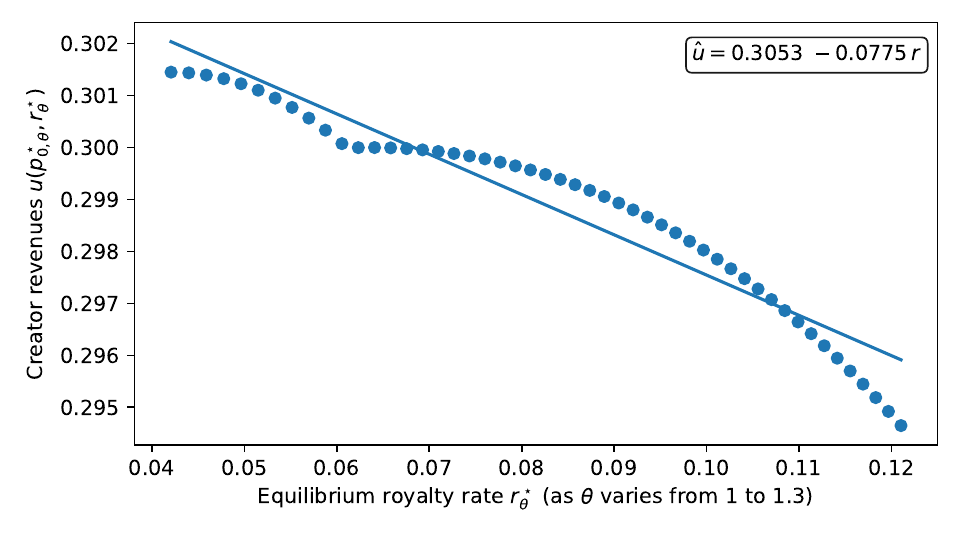}
    \caption{Realized revenue versus chosen royalties under overconfidence. Fix $b=1.2$ and $(a_0,a_1)=(0.5,0.05)$ with outside option $\Oopt(p_0)\defined a_0+(1+a_1)p_0$. For each $\theta\in[1,1.3]$, the overconfident creator chooses $(p_{0,\theta}^\star,r_\theta^\star)\in\arg\max_{p_0,r}\widehat u_c(p_0,r;\theta)$ under $\widehat F_\theta(v)=F(v/\theta)$, and the plot reports realized revenue $u(p_{0,\theta}^\star,r_\theta^\star)$ evaluated under the true distribution. The fitted regression line highlights the negative association between higher chosen royalties and lower realized revenue in this specification.}
    \label{fig:oc_u_vs_r_scatter}
\end{figure}

\begin{theorem}[Overconfidence raises royalties and lowers realized revenue]\label{thm:oc_uniform_full}
In the uniform example above with $p_0=0$, define the feasible set
\[
\mathcal{R}\defined \left\{ r\in[0,1): \frac{a_0}{1-r}\le b \right\} = \left[0,\, 1-\frac{a_0}{b}\right].
\]
Let $r^* \in \arg\max_{r\in\mathcal{R}} u(r;b)$ and $r(\theta) \in \arg\max_{r\in\mathcal{R}} \widehat u(r;b,\theta)$ for $\theta>1$.
Then:
\begin{enumerate}
\item \textbf{Royalties increase with overconfidence.} One has $r(\theta)> r^*$.
\item \textbf{Realized revenue falls under overconfidence.} One has $u(r(\theta);b)< u(r^*;b)$.
\end{enumerate}
\end{theorem}
In Figure~\ref{fig:oc_u_vs_r_scatter}, we show the negative correlation that emerges between revenues and equilibrium royalty rates chosen by overconfident creators. Our findings directionally align with empirical evidence presented in  \cite{Tunc2024NFTRoyalty}.

\begin{proof}[Proof of Theorem~\ref{thm:oc_uniform_full}]
Differentiate \cref{eq:oc_uniform_true_revenue_full} on $\mathcal{R}$:
\begin{align}
    {\frac{\partial u}{\partial r}}=\frac{1}{2}\left(b-\frac{a_0^2}{b}\cdot \frac{1+r}{(1-r)^3}\right),
\qquad
{\frac{\partial^2 u}{\partial r^2}}=-\frac{a_0^2}{b}\cdot \frac{2+r}{(1-r)^4}<0,
\end{align}
so $u(\cdot\,;b)$ is strictly concave and $r^*$ is unique. {We claim $r^*$ is interior. By strict concavity, we know $r^*$ must either be the lower boundary, the upper boundary, or an interior point satisfying the FOC. To prove the claim, it suffices to rule out the boundary optimality:
\begin{enumerate}[label = (\roman*)]
    \item Notice $\frac{\partial u}{\partial r}(0) = \frac{b^2 -a_0^2}{2b} > 0$. By continuity of the objective function, any small increase in $r$ from 0 will strictly improves the objective value. Thus, $r^*\neq 0$. 
    \item Notice $\frac{\partial u}{\partial r}\inparen{1-\frac{a_0}{b}} = -\frac{b^2 \left(1-\frac{a_0}{b}\right)}{a_0}$. Recall, $0< a_0 < b \implies 0<\frac{a_0}{b} < 1$, then $\frac{\partial u}{\partial r}\inparen{1-\frac{a_0}{b}}<0$. Similarly, any decrease from $1-\frac{a_0}{b}$ will improve the objective value, so that $r^*\neq 1-\frac{a_0}{b}$.
\end{enumerate}
Thus, we know $r^*$ must be interior that satisfies
\begin{align}
    b^2(1-r)^3 = a_0^2(1+r)
\end{align}

}
Similarly,
\begin{align}
    \frac{\partial \widehat u}{\partial r}
=
\frac{1}{2}\left(\theta b-\frac{a_0^2}{\theta b}\cdot \frac{1+r}{(1-r)^3}\right),
\qquad 
{\frac{\partial^2 \widehat u}{\partial r^2} = -\frac{a_0^2}{b\theta} \cdot \frac{2+r}{(1-r)^4} < 0}
\end{align}
{Notice$\frac{\partial \widehat u}{\partial r}(0) = \frac{b^2 \theta ^2-a_0^2}{2 b \theta } > 0$, so that lower boundary optimality can be excluded. However, $\frac{\partial \widehat u}{\partial r}(1-\frac{a_0}{b}) = \frac{b \left(a_0 \left(\theta ^2+1\right)-2 b\right)}{2 a_0 \theta }$, in which the sign depends on the strength of overconfidence $\theta$. Therefore, optimizing the perceived revenue over $\mathcal{R}$ could lead to either interior maximizer or upper boundary maximizer, depending on $\theta$. 
}
Nevertheless, any interior maximizer $r(\theta)$ satisfies
\begin{align}
    \theta^2 b^2 (1-r)^3 = a_0^2(1+r)
\end{align}
Define $\phi(r)\defined \frac{a_0^2(1+r)}{b^2(1-r)^3}$, which is strictly increasing on $\mathcal{R}$ {because $\frac{\partial \phi}{\partial r} = \frac{2 a_0^2 (r+2)}{b^2 (r-1)^4} > 0$}.
If both maximizers are interior, then $\phi(r^*)=1$ and $\phi(r(\theta))=\theta^2>1$. {Strict monotonicity of $\phi$ implies} $r(\theta)>r^*$.

If $r(\theta)$ is not interior, strict concavity of $\widehat u(\cdot\,;b,\theta)$ implies the maximizer is attained at {the upper} boundary point; {Recall $r^*$ is attained within the interior, that is $r^* \in \mathcal{ R}^\circ$, implying $r(\theta) > r^*$.}

Finally, {we claim $r(\theta) \neq r^*$. Recall, we've shown $r^* \in \mathcal{R}^\circ$ and $r(\theta)$ is either interior or the upper boundary point. They cannot equal if one is interior, and the other is the upper boundary point. If both are interior and $r^* = r(\theta)$, then we must have $\phi(r^*) = \phi(r(\theta)) \implies 1 = \theta^2 \implies \theta =1$, which cannot be true under the assumption $\theta >1$.
} Since $r^*$ uniquely maximizes $u(\cdot\,;b)$ over $\mathcal{R}$, we have $u(r(\theta);b)< u(r^*;b)$.
\end{proof}

\subsection{Robustness under Partial Willingness-to-Pay Extraction}\label{app:partial_extraction}

In the main analysis, we assume that speculators can fully extract the end buyer's realized willingness to pay, so that resale prices equal $V$. In practice, extraction is likely partial. Since partial extraction scales down resale revenues uniformly, one would expect it to affect results quantitatively without altering the main qualitative takeaways. Nonetheless, it is useful to verify this intuition; we do so by considering a reduced-form model in which speculators extract only a fraction $\phi\in (0,1)$ of the end buyer's valuation. Resale prices therefore equal $\phi V$, and royalties apply to $\phi V$ rather than $V$. 

\textbf{Benchmark Results:} In the benchmark setting of Section~\ref{sec:benchmark}, since the creator anticipates that the speculator will participate whenever expected profits are non-negative, the creator can extract the speculator's entire expected resale surplus by setting the initial price to $p_0 = \phi\E[V]$. In this case, the creator's profit with speculator is $\phi\E[V]$.

By contrast, in the absence of a speculator, the creator's optimal posted-price revenue equals $\max_{p_0} p_0\Pr[V\geq p_0]$. The excess profit generated by access to a speculator is therefore:
\begin{align}
    \Delta(\phi):= \phi\E[V] - \max_{p_0} p_0\Pr[V\geq p_0].
\end{align}

Assume $V$ is non-degenerate and non-negative, and has finite mean. Define the threshold:
\begin{align}
    \phi^*:=\frac{\max_{p_0\ge 0} p_0\Pr(V\ge p_0)}{\E[V]} \in (0,1) \label{eqn:wtpcritical}
\end{align}
Recall the Markov inequality is strict under non-degeneracy, then $\phi^*<1$. Also, $\phi^* >0$, because $\Pr[V\geq 0] > 0$ by our assumption and $p_0 =0$ cannot be optimal.
It follows that for any $\phi \in \inparen{\phi^*, 1}$,
\begin{align}
    \phi \E[V] > \phi^* \E[V] = \max_{p_0\ge 0} p_0\Pr(V\ge p_0),
\end{align}
and hence $\Delta(\phi)>0$. Thus, as long as willingness-to-pay extraction is sufficiently strong, the presence of an informed speculator strictly dominates.

\paragraph{Remark.}
The threshold $\phi^*$ is distribution-dependent and cannot be bounded away from one without additional assumptions. Nevertheless, for common distributions, $\phi^*$ is moderate. For example:
\begin{enumerate}[label = (\roman*)]
    \item If $V\sim \operatorname{Exp}(\lambda)$, then $\Pr[V \geq p_0] = \exp(-\lambda p_0)$ and $\E[V] = 1/\lambda$. Maximizing $p_0 \exp(-\lambda p_0)$ yields $p_0^* = \frac{1}{\lambda}$ and $\phi^* = \frac{1}{4} \cdot \frac{1}{\E[V]} = 1/4 * 2 = 1/2$.

    \item If $V\sim U[0,1]$, then $\Pr[V\geq p_0] = 1-p_0$ for $p_0 \in [0,1]$. Maximizing $p_0(1-p_0)$ yields $ p_0^* = 1/2$ and $\phi^*=1/2$.
\end{enumerate}

Now, We turn to Theorem~\ref{thm:riskneutral}. Under partial extraction, the speculator can only resell at $\phi V$, so $p_2 = \phi V$. The creator's problem therefore becomes:
\begin{align}
    \max_{p_0, r} &\quad p_0 + r\phi \E[V] - c\\
    \text{s.t.} &\quad \phi E[V] (1-r) - p_0 = 0. \label{eqn:wtpspecparticipation}
\end{align} 
Substituting the binding constraint \eqref{eqn:wtpspecparticipation} into the objective yields the constant payoff $\phi \E[V] -c$, independent of the royalty rate $r$. Hence, as in the baseline model, the creator is indifferent over the choice of $r$, and Theorem~\ref{thm:riskneutral} continues to hold under partial willingness-to-pay extraction. In fact, for thm~\ref{thm:riskneutral}, the results (no friction, royalty plays no role) hold for all $\phi\in (0,1)$. Notice creator always can use any $r$ to transfer the entirety of speculator revenue $\phi \E[V]$ for any $\phi$.

\textbf{Risk-Sharing:} In the setting of Section~\ref{sec:riskaversion}, the speculator now resells at price $p_2 = \phi V$. The profits of the creator and the speculator are given by:
\begin{align}
    v_c(p_0,r) &= p_0 + r p_2  - c = p_0 + r\phi V - c \\
    v_s(p_0,r) &= (1-r)p_2 - p_0 = (1-r)\phi V - p_0 
\end{align}
Let $\E[V] = \mu$ and $\Var[V] = \sigma^2$.
Assuming mean-variance utility, the agents' utilities are:
\begin{align}
    u_c(p_0,r) &= \E[v_c(p_0,r)] - \eta_c \Var[v_c(p_0,r)] = p_0 + r\phi\mu - c - \eta_c r^2\phi^2 \sigma^2 \\
    u_s(p_0,r) &= \E[v_s(p_0,r)] - \eta_s \Var[v_s(p_0,r)] = \inparen{1-r}\phi \mu - p_0 - \eta_s\inparen{1-r}^2\phi^2\sigma^2
\end{align}
The creator chooses the mint price $p_0$ and royalty rate $r$ to maximize $u_c$, subject to speculator's participation, $u_s \geq 0$. The problem is:
\begin{align}
    \max_{p_0, r} &\quad p_0 + r\phi\mu - c - \eta_c r^2\phi^2 \sigma^2\\
    \text{s.t.} &\quad \inparen{1-r}\phi \mu - p_0 - \eta_s\inparen{1-r}^2\phi^2\sigma^2 \geq 0 \label{eqn:riskspecparticipate}
\end{align}
At optimal, speculator participation constrain \eqref{eqn:riskspecparticipate} will bind, implying:
\begin{align}
    p_0 = \inparen{1-r}\phi \mu - \eta_s\inparen{1-r}^2\phi^2\sigma^2 \label{eqn:riskspecbind}
\end{align}

Now, we consider two cases:
\begin{enumerate}[label=(\roman*)]
    \item If the royalty option is removed ($r = 0$), then:
\begin{align}
    p_{0, r=0}^* = \phi \mu - \eta_s \phi^2 \sigma^2,\quad\text{and } u_{c, r=0}^* = \phi \mu - \eta_s \phi^2 \sigma^2 - c.
\end{align}

    \item If the royalty option is retained ($r>0$), plugging the binding constrain \eqref{eqn:riskspecparticipate} to the objective function gives:
\begin{align}
    \max_r \quad \phi\mu - c -\sigma ^2 \phi ^2 \left(r^2 \eta _c+(r-1)^2 \eta _s\right) \label{eqn:riskyesobj}.
\end{align}
The objective is concave in $r$ whenever both agents are weakly risk averse ($\eta_s, \eta_c \geq 0$), i.e.,
\begin{align}
    \frac{\partial^2 }{\partial^2 r} \inparen{\phi\mu - c -\sigma ^2 \phi ^2 \left(r^2 \eta _c+(r-1)^2 \eta _s\right)} = -2 \sigma ^2 \phi ^2 \eta _c-2 \sigma ^2 \phi ^2 \eta _s \geq 0,
\end{align}
FOC yields the unique optimal royalty rate:
\begin{align}
    0 = \frac{\partial }{\partial r} \inparen{\phi\mu - c -\sigma ^2 \phi ^2 \left(r^2 \eta _c+(r-1)^2 \eta _s\right)} 
    &= -2 \sigma ^2 \phi ^2 \left(r \eta _c+(r-1) \eta _s\right) \\
    \implies 
    r^* &= \frac{\eta _s}{\eta _c+\eta _s},
\end{align}
which coincides with the baseline result. Plugging $r^*$ back to the binding participation constraint gives the optimal price:
\begin{align}
    p_0^* = \phi \cdot \frac{\eta_c}{\inparen{\eta_c + \eta_s}^2} \cdot \inparen{\inparen{\eta_c + \eta_s}\mu - \phi \cdot \eta_c \eta_s \sigma^2}, 
\end{align}
which is non-negative provided $\mu \geq \phi \cdot \sigma^2 \frac{\eta_c \eta_s}{\eta_c + \eta_s}$. 

Under these conditions, the creator's optimal utility with royalties equals:
\begin{align}
    u_c^*
    = \phi \cdot \mu - c - \inparen{\frac{\eta_c \eta_s}{\eta_c + \eta_s}}\phi^2\sigma^2.
\end{align}
\end{enumerate}

To assess the value of royalties, we compare $u_c^*$ and $u_{c, r=0}^*$:
\begin{align}
    u_c^* - u_{c, r=0}^* 
    &= \phi \cdot \mu - c - \inparen{\frac{\eta_c \eta_s}{\eta_c + \eta_s}}\phi^2\sigma^2  - \inparen{\phi \mu - \eta_s \phi^2\sigma^2 -c} \\
    &= \sigma ^2 \phi ^2 \inparen{\frac{ \eta _s^2}{\eta _c+\eta _s}}
\end{align}
which is positive whenever both agents are risk averse ($\eta_c, \eta_s >0$). Thus, even under partial willingness-to-pay extraction, the option to use royalties strictly improves the creator's utility relative to a pure upfront-price mechanism. 

\textbf{Information Asymmetry:} As in the setting of Section~\ref{sec:information_asymmetry}, suppose the speculator observes the realization $v$ of $V$ before anyone else. But due to limited surplus extraction, the 
speculator can only resell at $p_2 = \phi v$ for some $\phi \in (0,1)$.

If no royalty is imposed, trade between the creator and the speculator occurs if and only if:
\begin{align}
    \phi v - p_0 \geq 0 \iff p_0 \leq \phi v \iff v \geq \frac{p_0}{\phi}.
\end{align}

The creator's expected profit is:
\begin{align}
    p_0 \cdot \E[\Ind_{V\geq p_0/\phi}] - c = p_0 \Pr\inbrak{V \geq \frac{p_0}{\phi}} -c.
\end{align}
Simplifying out the constant cost $c$, the creator solves:
\begin{align}
    \max_{p_0} \quad p_0 \Pr\inbrak{V \geq \frac{p_0}{\phi}}
\end{align}
By Markov's inequality,
\begin{align}
    \frac{p_0}{\phi} \Pr\inbrak{V \geq \frac{p_0}{\phi}} \leq \E[V] \implies p_0 \Pr\inbrak{V \geq \frac{p_0}{\phi}} \leq \phi E[V] < \E[V], \quad \forall p_0 \geq 0
\end{align}
Recall, the inequality is strict for any non-degenerate distribution.

Thus, when the speculator is better informed and no royalties are used, the creator cannot fully exploit the speculator's information advantage and cannot extract the entire expected surplus $\E[V]$.

We now consider the role of royalties under information asymmetry and partial willingness-to-pay.

At time 0, the creator sets an upfront price $p_0$ and a royalty rate $r$.
At time 1, the speculator observes the realized valuation $v$. Due to limited surplus extraction, the speculator can resell the asset to the end buyer at price $\phi v$, so that his net profit after royalty payments is $(1-r)\phi v - p_0$. The speculator therefore purchases the asset if and only if
\begin{align}
    (1-r)\cdot \phi v - p_0 \geq 0,\quad \text{or } (1-r)\cdot \phi v \geq p_0
\end{align}

If the speculator purchases the asset, the creator receives the upfront payment $p_0$ and royalty revenue $r\phi v$ later. Since the creator does not observe $v$, expected revenue in this case is
\begin{align}
    p_0 + r \E\inbrak{\phi V \mid (1-r)\cdot \phi V \geq p_0} \quad\text{w.p. }\Pr[(1-r)\cdot \phi V \geq p_0]
\end{align}
If the speculator does not purchase the asset, the creator may still sell directly to the end buyer. This occurs if and only if both the speculator declines and the end buyer's valuation exceeds the posted price, i.e.,
\begin{align}
    \inset{(1-r)\cdot \phi V \leq p_0} \cap \inset{V \geq p_0},
\end{align}
in which case the creator receives revenue $p_0$. 

Combining both cases, the creator's expected profit is
\begin{align}
    u_c &= \inparen{p_0 + r \E\inbrak{\phi V \mid (1-r)\cdot \phi V \geq p_0}}\Pr[(1-r)\cdot \phi V \geq p_0]\\ 
    &+ p_0 \Pr[\inset{(1-r)\cdot \phi V \leq p_0} \cap \inset{V \geq p_0}] - c
\end{align}

Simplifying, the creator's objective can be written as:
\begin{align}
    p_0(1-F(p_0)) + r \phi \int_{\frac{p_0}{\phi(1-r)}}^\infty v f(v)dv 
\end{align}

Hence, the creator solves
\begin{align}
    \max_{p_0, r}\quad p_0(1-F(p_0)) + r \phi \int_{\frac{p_0}{\phi(1-r)}}^\infty v f(v)dv - c \label{eqn:infoasymwtpobj}
\end{align}
Applying Leibniz's integral rule yields first-order conditions with respect to $p_0$ and $r$:
\begin{align}
    \frac{\partial}{\partial p_0}\inparen{p_0(1-F(p_0)) + r \phi\int_{\frac{p_0}{\phi(1-r)}}^\infty v f(v)dv} 
    &= 1-F(p_0) - p_0f(p_0) - r \frac{p_0 f\inparen{\frac{p_0}{\phi(1-r)}}}{\phi(1-r)^2}\\
    \frac{\partial}{\partial r}\inparen{p_0(1-F(p_0)) + r \phi \int_{\frac{p_0}{\phi(1-r)}}^\infty v f(v)dv}
    &= \phi \int_{\frac{p_0}{\phi(1-r)}}^\infty v f(v)dv - r \frac{p_0^2 f\inparen{\frac{p_0}{\phi(1-r)}}}{\phi\inparen{1-r}^3}
\end{align}

These expressions are algebraically involved and do not admit a tractable closed-form characterization under partial willingness-to-pay. Rather than solving for an interior solution, we focus on boundary optimality, which is sufficient for our purposes. 

We resume from the objection function \eqref{eqn:infoasymwtpobj}:
\begin{align}
    &p_0(1-F(p_0)) + r \phi \int_{\frac{p_0}{\phi(1-r)}}^\infty v f(v)dv\\ 
    &= p_0 \Pr[V> p_0] + r \E\inbrak{\phi V \mid (1-r)\phi V \geq p_0} \Pr[(1-r)\phi V \geq p_0]
\end{align}
Notice at $p_0 =0, r=1$, the creator's revenue is $\phi \E[V]$, which is the maximum amount of revenue that could be achieved. To see this, note that the creator can generate revenue only by either selling to the speculator or selling directly to the end buyer:
\begin{enumerate}[label = (\roman*)]
    \item If creator sells the product to speculator, then due to limited extraction of end buyer's willingness-to-pay, speculator can only charge $\phi v$, so a full royalty extracts all resale surplus in expectation.
    \item If creator sells to the end buyer directly, then consider:
    \begin{align}
        \Pr[\inset{(1-r)\cdot \phi V \leq p_0} \cap \inset{V \geq p_0}] &\leq \Pr[V\geq p_0]\\
        \implies
        p_0\Pr[\inset{(1-r)\cdot \phi V \leq p_0} \cap \inset{V \geq p_0}] &\leq p_0\Pr[V\geq p_0] \leq \E[V],
    \end{align}
    with strict inequality for non-degenerate distributions, implying that direct sales cannot exceed $\phi\E[V]$ when $\phi$ is sufficiently large, i.e., $\phi \in (\phi^*, 1)$ for threshold $\phi^*$ in \eqref{eqn:wtpcritical}.
\end{enumerate}

As shown earlier, mechanisms without royalties ($r=0$) strictly underperform this bound. Therefore, any optimal mechanism must involve a strictly positive royalty rate. While uniqueness of the optimal $(p_0,r)$ pair can be established under full willingness-to-pay, such a result is not required here. The boundary solution $(p_0=0, r=1)$ achieves the maximal feasible revenue $\phi\E[V]$ and suffices to establish optimality under partial willingness-to-pay.

The spirit of the analysis extends naturally to the multiunit case, though it is much more involved there, as can be seen from the existing proofs. There as well, the high-level takeaway is that the more capable speculators are at extracting buyer WTP, the more useful royalties are for creators.

\subsection{Robustness under liquidity effects}\label{app:liquidity}

Our baseline model isolates the surplus-extraction benefit of royalties, abstracting from any adverse effect they may have on secondary-market liquidity. As discussed in Section~\ref{sec:lit}, \cite{Tunc2024NFTRoyalty} document a negative correlation between royalty rates and creator revenues among NFTs that have already been listed. While we argue that this finding reflects a different object of analysis---\emph{ex post} outcomes conditional on listing, rather than the \emph{ex ante} participation decision we model---the possibility that high royalty rates discourage secondary-market activity is economically plausible and worth formalizing. In particular, if buyers shy away from NFTs with high royalty rates, then the benefit of surplus extraction must be weighed against the cost of reduced trade. 

To explore this, we take the adverse liquidity effect as given and incorporate it into our model in a reduced-form way. Suppose that a resale buyer is found with probability $(1-\theta r)$, where $\theta\in (0,1)$ and $r$ is the royalty rate. Thus, as the royalty rate increases, the probability of finding a buyer decreases (tempered by the scaling parameter $\theta$).

Conditional on finding a buyer, the speculator resells the asset at price $V$; otherwise, resale revenue is zero. Hence, $p_2 = V \cdot \Ind_{\text{buyer found}}$. 

The creator's expected profit is given by:
\begin{align}
    \E[v_c] &= p_0 + r \E[p_2] - c \\
    &= p_0 + r \E[V]\Pr(\text{buyer found}) - c\\
    &= p_0 + r \mu \inparen{1-\theta r} - c
\end{align}
The speculator's expected profit is
\begin{align}
    \E[v_s] = (1-r)\mu(1-\theta r) - p_0 
\end{align}
In equilibrium, the speculator's participation constraint binds. 
Substituting into the creator's objective yields the following linear program:
\begin{align}
    \max_{r\in[0,1]} &\quad -c+\mu -\theta  \mu  r \implies r^* = 0
\end{align}
Intuitively, royalties here act as a tax on resale that reduces the likelihood of trade. Because royalties do not raise the value created by a successful resale, the loss from fewer transactions strictly outweighs any additional royalty revenue, rendering non-zero royalties suboptimal.

We now extend the analysis to an information-asymmetry environment. 

Recall the timeline, at time 0, the creator decides price $p_0$ and royalty rate $r$. At time 1, the better informed speculator observes the realized valuation $v$ of $V$ prior to purchase. However, resale is uncertain: with probability $1-\theta r$ a buyer is found, while with probability $\theta r$ resale fails. Therefore, in expectation, the speculator purchase the asset if and only if
\begin{align}
    (1-r) \cdot (1-\theta r) v - p_0 \geq 0,\quad\text{or } (1-r) \cdot (1-\theta r) v \geq p_0
\end{align}

If the speculator purchases, the creator receives the upfront payment $p_0$ and subsequent royalty revenue $rp_2$. Due to information asymmetry, creator does not observe $v$, so she only knows $p_2 = V \cdot \Ind_{\text{buyer found}}$, then her expected revenue in this case is
\begin{align}
    &p_0 + r \E\inbrak{V \cdot \Ind_{\text{buyer found}} \mid (1-r) \cdot (1-\theta r) V \geq p_0} \\
    &= p_0 + r (1-\theta r)\E\inbrak{V \mid (1-r) \cdot (1-\theta r) V \geq p_0}
\end{align}
with probability $\Pr[(1-r) \cdot (1-\theta r) V \geq p_0]$.

If the speculator does not purchase, the creator may still profit $p_0$ by selling directly to the end buyer if speculator declines sale and the end buyer's valuation exceeds the posted price, i.e.,
\begin{align}
    \inset{(1-r) \cdot (1-\theta r) V < p_0} \cap \inset{V \geq p_0}
\end{align}

Therefore, the creator's expected profit is
\begin{align} 
    u_c &= \inparen{p_0 + r (1-\theta r)\E\inbrak{V \mid (1-r) \cdot (1-\theta r) V \geq p_0}} \Pr[(1-r) \cdot (1-\theta r) V \geq p_0]\\
    &+ p_0 \Pr\inbrak{\inset{(1-r) \cdot (1-\theta r) V < p_0} \cap \inset{V \geq p_0}} - c 
\end{align}

We claim under the adversarial effect of royalty, no royalty $(r=0)$ is optimal. The maximization problem $\max_{p_0, r} u_c$ is analytically intractable without additional
assumptions on the distribution of $V$. We therefore proceed via backward induction on speculator's purchasing decision to show boundary solutions involving $r=0$ is optimal: 

\begin{enumerate}[label = (\roman*)]
    \item Suppose speculator buys from the creator, which happens if and only if $(1-r)p_2 \geq p_0$.
    
    From the creator's information set, the speculator's maximum feasible resale revenue is $p_2 = V \cdot \Ind_{\text{buyer found}}$. Taking expectations over resale uncertainty, the expected resale revenue equals $(1-\theta r) \E[V]$. The creator solves
    \begin{align}
        \max_{p_0, r} &\quad p_0 + r(1-\theta r) \E[V]\\
        \text{s.t.} &\quad (1-r)(1-\theta r) \E[V] \geq p_0
    \end{align}
    Since the constrain will bind, the objective function can be further simplified
    \begin{align}
        \max_{r} (1-\theta r) \E[V] \implies r^* = 0
    \end{align}
    which results in an optimal revenue of $\E[V]$.

    \item Suppose speculator does not buy from the creator. Then, either creator succeeds in selling to the end buyer or no trades with zero revenue. The event that creator skips speculator and sells to the end buyer is equivalent to
    \begin{align}
        \inset{(1-r) \cdot (1-\theta r) V < p_0} \cap \inset{V \geq p_0},
    \end{align}
    which is a subset of event $\inset{V \geq p_0}$. 
    In this case, the expected revenue can be upper bounded by Markov's inequality:
    \begin{align}
         p_0 \Pr\inbrak{\inset{(1-r) \cdot (1-\theta r) V < p_0} \cap \inset{V \geq p_0}} \leq p_0 \Pr[V\geq p_0] \leq \E[V]
    \end{align}
    In particular, the last inequality is strict for non-degenerate distribution. Under this scenario, the optimal revenue of direct sale is strictly less than $\E[V]$ for all $r$.
\end{enumerate}
Since we have exhausted all the cases, and the creator attains maximally possible revenue $\E[V]$ only when $r=0$, it follows that the overall optimum must satisfy $r^*=0$.

\clearpage
\pagenumbering{arabic}     
\setcounter{page}{1}

\renewcommand{\thesection}{OA\arabic{section}}  
\setcounter{section}{0}
\renewcommand{\thesubsection}{\thesection.\arabic{subsection}}

\begin{center}
    \Large{Online Appendix (OA) For}
\end{center}

\begin{center}
    \Large{Economics of NFTs: The Value of Creator Royalties}
\end{center}

\begin{center}
    {(Author names blinded for peer review)}
\end{center}

\noindent This online appendix contains the proofs for the manuscript.

\begin{proof}[\bf Proof of Lemma~\ref{lem:optpbase}]

Note that $\Pr[V>p_0] = 1-F(p_0)$, with $f,F$ being the density and cumulative distribution functions of $V$. Take the first and second-order derivatives of the objective function:
\begin{align}
    FOC \quad \Leftrightarrow & \quad \frac{\partial}{\partial p_0} \Big (p_0 (1-F(p_0)) \Big ) = 1-F(p_0) - p_0 f(p_0) = 0 \label{eq:foc}\\
    SOC \ Concavity \quad  \Leftrightarrow & \quad \frac{\partial^2}{\partial p_0^2} \Big (p_0 (1-F(p_0) \Big ) = -2f(p_0) - p_0 f'(p_0) \le 0. \label{eq:soc}
    \end{align}
From \eqref{eq:soc}, concavity holds iff $p_0 \geq -2 \frac{f(p_0)}{f'(p_0)}$. For example, this trivially holds for the uniform distribution given $f\geq0$, $f'=0$ and $p_0 \geq 0$.

Concavity is too strict of a requirement for some other distributions, but in general, quasi-concavity, or at least uni-modality, suffice and will  hold. In such cases, the first-order condition \eqref{eq:foc} implies that $p_0^*$ is the solution to:
\begin{align*}
    p_0 = \frac{1-F(p_0)}{f(p_0)}.
\end{align*}
The right side can be interpreted as the inverse hazard rate (failure rate). A few examples: When $V$ is uniform between $[v_L,v_H]$, $p_0^*=\frac{v_H }{2}$. When $V$ is exponential with parameter $\lambda$, the objective is unimodal and $p_0^* = \frac{1}{\lambda}$. When $V$ is normally distributed with parameters $\mu, \sigma$, the objective is unimodal and there is still a unique solution for $p_0^*$ but it is not in closed-form.

\end{proof}

\begin{proof}[\bf Proof of Corollary~\ref{cor:nono}]
We sketch out how Markov's inequality leads to the desired result with an example. 

\begin{example}[Markov's Inequality at Work]\label{eqn:2pointvalue}
As an example, define the two-point distribution $V \sim v_L$ with prob. 1/2 and $v_H$ with prob. 1/2, where $0 < v_L < v_H$.
Then $\E[V] = \frac{v_L+v_H}{2}$, but the creator's revenues are given by
    $\max_{p_0} p_0 \cdot \Pr [ V \ge p_0 ] = \max\inparen{ v_L, \frac{v_H}{2} }$,
which is strictly less than $\E[V]$, given $v_L > 0$.
\end{example}

\begin{table}[ht]
    \centering
    \begin{tabular}{@{}cccccc@{}} 
        \toprule 
        \textbf{Price range} & \textbf{Best $p_0$} & $\bf \Pr(V \geq p_0)$ & \textbf{Creator E[Rev]} & \textbf{Spec. E[Rev]} & \textbf{Buyer E[Rev]} \\
        \midrule 
        $p_0 \leq v_L$ & $v_L$ & 1 & $v_L$ & absent & $(v_H - v_L)/2$ \\
        $v_L < p_0 \leq v_H$ & $v_H$ & $1/2$ & $v_H/2$ & absent & 0 \\
        $v_H < p_0$ & indiff. & 0 & 0 & absent & 0 \\
        \bottomrule 
    \end{tabular}
    \caption{Creator and end-buyer revenues from Example~\ref{eqn:2pointvalue} without the speculator or royalties.}
    \label{tab:nono}
\end{table}

\noindent To understand Example~\ref{eqn:2pointvalue}, consider the possible ranges for $p_0$ given in Table~\ref{tab:nono}. If $p_0 \leq v_L$, the probability of trade is 1 and the optimal $p_0=v_L$, which leads to expected revenues of $v_L \cdot 1$. If $p_0 > v_H$, the probability of trade is 0 and there are no revenues to be made. And if $p_0$ is between $[v_L, v_H]$, the probability of trade is 1/2. In the latter case, the optimal $p_0=v_H$ and expected revenues are $v_H \cdot 1/2$.
\end{proof}

\begin{proof}[\bf Proof of Corollary~\ref{cor:noyes}]
    The proof is outlined in the text preceding the corollary.
\end{proof}

\begin{proof}[\bf Proof of Theorem~\ref{thm:riskneutral}]
    When both the creator and the speculator are risk neutral, the creator's maximization problem becomes
    \begin{align*}
    \max_{p_0, r} \quad E\left [ v_c \right]\\ 
    \mbox{subject to:} \quad \mu(1-r) -p_0 = 0, 
    \end{align*}
    with $v_c = p_0 + r p_2  - c$, and $p_2=V$. Taking the expectation, $E[v_c] = p_0 + r\mu - c$, the problem becomes
    \begin{align*}
    \max_{r \in [0,1]} \quad \mu (1-r) + r \mu - c , 
    \end{align*}
    which simplifies to $\max_r (\mu - c ) = \mu - c $.
    The above is conditional on the creator preferring to sell the NFT to the speculator at time 1, rather than waiting with the hope of selling  the NFT to the end-buyer at time $2$.

    This follows from the Markov inequality argument in Corollary~\ref{eqn:2pointvalue} (after noting $\mu-c = E[V] - c $). Intuitively, the creator prefers to sell the NFT to the speculator to piggyback on of the speculator's access to the aftermarket.
\end{proof}

\begin{proof}[\bf Proof of Lemma~\ref{lem:risk}]

The objective function in \eqref{eqn:creator_utility} can be written as
\begin{equation*}
    E\left [ v_s \right] - \eta_c \Var_s[v_s] = p_0 + r \mu - c - \eta_c r^2 \sigma^2.
\end{equation*}
Given Equation~\ref{eqn:speculator_profits} will bind, and re-arranging terms, the creator's maximization problem can be rewritten as a quadratic problem over $r$:
\begin{align}
\max_{r} \quad p_0 + r \mu - c - \eta_c r^2 \sigma^2 \label{eqn:optwithr}\\
\mbox{subject to:} \quad p_0 = \mu(1-r)-\eta_s \sigma^2(1-r)^2. \label{eqn:rest}
\end{align}

\subsubsection*{Solution Without Royalties ($r=0$)}
If the option of royalties is removed entirely ($r=0$), as some exchanges are contemplating doing, then the creator's optimization problem simplifies. From Equation~\ref{eqn:rest}, the price at which the speculator is willing to purchase the NFT is given by 
\begin{equation*}
    p_{0,r=0}^* = \mu - \eta_s \sigma^2.
\end{equation*} Replacing this in the objective in \eqref{eqn:optwithr} while setting $r=0$, leads to a creator utility absent royalties given by:
\begin{equation}
    \label{eqn:NR}
    u_{c,r=0}^* = \mu - c  - \eta_s \sigma^2.
\end{equation}

\subsubsection*{Solution With Royalties ($r>0$)}
Replacing the constraint \eqref{eqn:rest} in the objective function simplifies the problem to a quadratic over $r$:
\begin{equation}
    \label{eqn:optafterconstraint}
    \max_{r} \mu - c - (\eta_s(1-r)^2 + \eta_cr^2)\sigma^2.
\end{equation}
Solving gives the optimal royalty rate, $r$ as follows:
\begin{align*}
	r^*= \frac{\eta_s}{\eta_s + \eta_c}.
\end{align*}
Replacing this in \eqref{eqn:rest} gives an optimal price of
\begin{align}
    p_0^* = \frac{\eta_c}{(\eta_c + \eta_s)^2} \Big ( (\eta_c + \eta_s) \mu - \eta_c \eta_s \sigma^2 \Big ).
\end{align}
For the price to be valid, it must be non-negative, implying
\begin{equation}
    \mu  \ge  \sigma^2 \frac{\eta_c \eta_s}{\eta_c + \eta_s}.
\end{equation}

Plugging these into the utility Equation~\ref{eqn:optafterconstraint} gives
\begin{align*}
    \mu - c - (\eta_s(1-{r^*})^2 + \eta_c{r^*}^2)\sigma^2 &= \mu - c - \inparen{ \eta_s \inparen{ \frac{-\eta_c}{\eta_c+\eta_s} }^2 + \eta_c \inparen{\frac{\eta_s}{\eta_c + \eta_s} }^2 } \sigma^2 \\
    &= \mu - c - \inparen{\frac{\eta_c \eta_s}{\eta_c + \eta_s}} \sigma^2.
\end{align*}
This is valid as long as it's non-negative, implying
\begin{equation}
    \mu - c  \ge  \sigma^2 \frac{\eta_c \eta_s}{\eta_c + \eta_s}.
\end{equation}
Thus, if the above is satisfied, it guarantees the price is also non-negative.

Finally, we need to consider the speculator's participation constraint $u_s \ge 0$. Plugging the results into $u_s(p_0, r) = (1-r)\mu -p_0 -\eta_s (1-r)^2\sigma^2$, we obtain $u_s^*  = 0$, confirming the speculator participates.
\end{proof}

\begin{proof}[\bf Proof of Theorem~\ref{thm:royalhigh}]
The creator profit with royalties is given in Equation~\ref{eqn:optr}, and the creator revenue without royalties is given in Equation~\ref{eqn:NR}.  Thus we must show that

\begin{equation}
    \mu - c - \inparen{\frac{\eta_c \eta_s}{\eta_c + \eta_s}} \sigma^2 \ge     \mu - c  - \eta_s \sigma^2.
\end{equation}

The difference is 
\begin{align}
    \sigma^2 \inparen{ \eta_s - \frac{\eta_c \eta_s}{\eta_c + \eta_s} } &= \sigma^2 \inparen{ \frac{\eta_c \eta_s + \eta_s^2 - \eta_c \eta_s}{\eta_c + \eta_s}} \\
    &= \sigma^2 \inparen{ \frac{\eta_s^2}{\eta_c + \eta_s}}.
\end{align}

which is positive whenever $\eta_c$ and $\eta_s$ are both positive.
Thus the option of royalties leads to strictly higher utility for the seller whenever $\eta_s > 0$ and $\eta_c + \eta_s > 0$. 

{\bf Comparative statics:} Denote the difference in utilities between royalties and no-royalties as a function $f$ in three variables, namely $\eta_s, \eta_c, \sigma$, that is
\begin{equation}
    f(\eta_s, \eta_c,\sigma)= \frac{\eta_s^2 \sigma^2}{\eta_s+\eta_c}.
\end{equation} 
It is straightforward to check the first derivatives that if $\eta_s,\eta_c > 0$, then $f(\eta_s,\eta_c,\sigma)$ is increasing in $\eta_s$ and $\sigma$, but decreasing in $\eta_c$. 

\end{proof}

\begin{proof}[\bf Proof of Corollary~\ref{cor:TRrisk}]
    The proof is outlined in the text preceding the corollary and the region of interest is computed numerically.
\end{proof}

\begin{proof}[\bf Proof of Corollary~\ref{cor:infoab}]
    The proof is outlined in the text preceding the corollary.
\end{proof}

\begin{proof}[\bf Proof of Lemma~\ref{lem:optinfoasym}]

As a first step, we can rewrite the creator's objective function \eqref{eq:optinfoasymu}. Ignoring the constant $c$, we have
\begin{align*}
    &\Big ( p_0 + r E[V|(1-r)V \ge p_0] \Big ) \cdot \Pr [ (1-r)V\geq p_0] +p_0 \Pr [ (1-r)\cdot V < p_0 \ \cap \ V \ge p_0] \\
    &= \inparen{ p_0 + r E\inbrak{ V \suchthat (1-r)V \ge p_0 } }\inparen{ 1-F\inparen{\frac{p_0}{1-r}}} + p_0 \inparen{ F\inparen{\frac{p_0}{1-r}} - F(p_0)} \\
     &= \inparen{ p_0 + r \cdot \frac{ \int_{\frac{p_0}{1-r}}^\infty v f(v) dv }{1 - F\inparen{\frac{p_0}{1-r} } } }\inparen{ 1-F\inparen{\frac{p_0}{1-r}}} + p_0 \inparen{ F\inparen{\frac{p_0}{1-r}} - F(p_0)} \\
     &= p_0 \inparen{ 1- F \inparen{\frac{p_0}{1-r} } } + r \int_{\frac{p_0}{1-r}}^\infty v f(v) dv + p_0 \inparen{ F\inparen{\frac{p_0}{1-r}} - F(p_0)} 
     \\
     &= p_0 (1 - F(p_0)) + r \int_{\frac{p_0}{1-r}}^\infty v f(v) dv
\end{align*}

Thus the creator's maximization problem is:
\begin{equation*}
    \max_{p_0,r} p_0 (1 - F(p_0)) + r \int_{\frac{p_0}{1-r}}^\infty v f(v) dv - c.
\end{equation*}

To find the first-order conditions on $p_0$, we can use Leibniz's integral rule to obtain:
\begin{equation*}
    \frac{\partial}{\partial p_0} \inparen{ p_0 (1 - F(p_0)) + r \int_{\frac{p_0}{1-r}}^\infty v f(v) dv } = 1-F(p_0) - p_0 f(p_0) - r\frac{p_0 f\inparen{ \frac{p_0}{1-r} } }{(1-r)^2}
\end{equation*}
and
\begin{equation*}
    \frac{\partial}{\partial r} \inparen{ p_0 (1 - F(p_0)) + r \int_{\frac{p_0}{1-r}}^\infty v f(v) dv } = \int_{\frac{p_0}{1-r}}^\infty v f(v) dv - \frac{rp_0^2 f\inparen{\frac{p_0}{1-r}}}{(1-r)^3}.
\end{equation*}
\end{proof}

\begin{proof}[\bf Proof of Theorem~\ref{thm:asym}]

Starting with the expected creator revenue in Equation~\ref{eq:optinfoasymu}, the creator's optimization problem is written as
\begin{equation}
    \max_{p_0, r} \quad \inparen{p_0 + r E\inbrak{V\suchthat (1-r)V\ge p_0 }}\cdot \Pr [ (1-r)V \geq p_0] + p_0 \Pr [ (1-r)\cdot V  < p_0 \cap  V \ge p_0]
\end{equation}
Notice that
\begin{equation}
    \Pr [ (1-r)\cdot V < p_0 \cap  V \ge p_0] =  \Pr[(1-r)V \le p_0] - \Pr[ V \le p_0 ],
\end{equation}
because 
\begin{align*}
    \Pr [ (1-r)\cdot V  < p_0 \cap  V \ge p_0] 
    &= \Pr[\{(1-r)\cdot V  < p_0\} \cap \{V\geq p_0\} ] \\
    &= \Pr[\{V<\frac{p_0}{1-r}\} \cap \{V\geq p_0\}] \\
    &= \Pr[V\leq \frac{p_0}{1-r}] - \Pr[V\leq p_0] \\
    &= \Pr[(1-r)V \le p_0] - \Pr[ V \le p_0 ].
\end{align*}
From there we use the above identity to simplify the objective function:
\begin{align*}
    &\inparen{p_0 + r E\inbrak{V\suchthat (1-r)V\ge p_0 }}\cdot \Pr [ (1-r)V \geq p_0] + p_0 \Pr [ (1-r)\cdot V  < p_0 \cap  V \ge p_0] \\ 
    &= \inparen{p_0 + r E\inbrak{V\suchthat (1-r)V\ge p_0 }}\cdot \Pr [ (1-r)V \geq p_0] + p_0 \inparen{\Pr[(1-r)V \le p_0] - \Pr[ V \le p_0 ]} \\
    &= \inparen{p_0 + r E\inbrak{V\suchthat (1-r)V\ge p_0 }}\cdot \Pr [ (1-r)V \geq p_0] + p_0 \cdot \Pr[(1-r)V \le p_0] - p_0 \cdot \Pr[ V \le p_0 ] \\
    &= \inparen{p_0 + r E\inbrak{V\suchthat (1-r)V\ge p_0 }}\cdot \Pr [ (1-r)V \geq p_0] + p_0 \cdot \inparen{1-\Pr[(1-r)V \ge p_0]} - p_0 \cdot \Pr[ V \le p_0 ] \\
    &= \inparen{p_0 + r E\inbrak{V\suchthat (1-r)V\ge p_0 }}\cdot \Pr [ (1-r)V \geq p_0] + p_0 -p_0 \cdot \Pr[(1-r)V \ge p_0]- p_0 \cdot \Pr[ V \le p_0 ] \\
    &= \inparen{r E\inbrak{V\suchthat (1-r)V\ge p_0 }}\cdot \Pr [ (1-r)V \geq p_0] + p_0- p_0 \cdot \Pr[ V \le p_0 ] \\
    &= \inparen{r E\inbrak{V\suchthat (1-r)V\ge p_0 }}\cdot \Pr [ (1-r)V \geq p_0] + p_0\cdot \inparen{1-\Pr[ V \le p_0 ]}
\end{align*}

which yields a simplified objective function:
\begin{equation}\label{eqn:altuinfoasym}
    p_0 \cdot \Pr[V> p_0] + r\cdot E\inbrak{V\suchthat (1-r)V\ge p_0 }\cdot \Pr [ (1-r)V \geq p_0].
\end{equation}

Now, notice that setting $p_0 = 0$, and $r = 1$, the creator earns revenue $E[V]$, which is the maximum amount of revenue that could be achieved. For this solution to be unique, we need to verify what happens if the creator chooses an $r<1$.

When $r < 1$, the creator's revenue is

\begin{align}
    p_0 \int_{p_0}^\infty f_V(v) dv + r \int_{\frac{p_0}{1-r}}^\infty v f_V(v) dv &= p_0 \int_{p_0}^{\frac{p_0}{1-r}} f_V(v) dv + \int_{\frac{p_0}{1-r}}^\infty (rv + p_0) f_V(v) dv \label{eqn:le1} \nonumber\\
    &\le p_0 \int_{p_0}^{\frac{p_0}{1-r}} f_V(v) dv + \int_{\frac{p_0}{1-r}}^\infty v f_V(v) dv \nonumber \\
    & \le \int_{p_0}^\infty v f_V(v) dv  \\
    &= E[V] - \int_{0}^{p_0} vf_V(v) dv. \label{eqn:le3}
\end{align}

Now, consider three cases $\Pr[ V < p_0 ] > 0$, $\Pr\inbrak{ p_0 < V < \frac{p_0}{1-r} } > 0$  and $\Pr[ V > p_0] > 0$.

\textbf{Case 1:} If $\Pr[ V < p_0 ] > 0$, then 
\begin{equation*}
    \int_0^{p_0} v f_V(v) dv > 0.
\end{equation*}
By Equation~\ref{eqn:le3}, the creator revenue is bounded by 
\begin{equation*}
    E[V] - \int_{0}^{p_0} vf_V(v) dv  < E[V].
\end{equation*}

\textbf{Case 2:} If $\Pr\inbrak{ p_0 < V < \frac{p_0}{1-r} }$, then 
\begin{equation*}
    p_0 \int_{p_0}^{\frac{p_0}{1-r}} f_V(v) dv < \int_{p_0}^{\frac{p_0}{1-r}} v f_V(v) dv. 
\end{equation*}
So by Equation~\ref{eqn:le1}, the creator revenue is bounded by 
\begin{align*}
    p_0 \int_{p_0}^{\frac{p_0}{1-r}} f_V(v) dv + \int_{\frac{p_0}{1-r}}^\infty (rv + p_0) f_V(v) dv &< \int_{p_0}^{\frac{p_0}{1-r}} vf_V(v) dv + \int_{\frac{p_0}{1-r}}^\infty (rv + p_0) f_V(v) dv \\
    &\le \int_{p_0}^\infty v f_V(v) dv \\
    &\le E[V].
\end{align*}

\textbf{Case 3:}
If $\Pr\inbrak{ V > \frac{p_0}{1-r}}$, then

\begin{equation*}
    \int_{\frac{p_0}{1-r}}^\infty (rv + p_0) f_V(v) dv < \int_{\frac{p_0}{1-r}}^\infty v f_V(v) dv.
\end{equation*}

So by Equation~\ref{eqn:le1}, the creator's revenue is bounded by 

\begin{align*}
    p_0 \int_{p_0}^{\frac{p_0}{1-r}} f_V(v) dv + \int_{\frac{p_0}{1-r}}^\infty (rv + p_0) f_V(v) dv &< p_0 \int_{p_0}^{\frac{p_0}{1-r}} f_V(v) dv + \int_{\frac{p_0}{1-r}}^\infty v f_V(v) dv \\
    &\le \int_{p_0}^\infty v f_V(v) dv \\
    &\le E[V].
\end{align*}

Therefore, in all three situations, the creator \emph{cannot} achieve revenue $E[V]$.  Thus the \emph{unique} optimal solution is $r = 1, p_0 = 0$.

\end{proof}

\begin{proof}[\bf Proof of Lemma~\ref{cor:asym_exponential}]
    First, we show that if $V$ is exponentially distributed with parameter $\lambda$, then
    \begin{equation}
        \label{eqn:expprob}
        \Pr[V(1-r) > p_0 ] = e^{-\lambda \frac{p_0}{1-r}}
    \end{equation}
    and
    \begin{equation}
        \label{eqn:expcondexp}
        E\inbrak{V\suchthat V(1-r)>p_0} = \frac{p_0}{1-r} + E[V].
    \end{equation}

    To see this, consider that $V$ exponentially distributed implies
    \begin{equation*}
        \Pr[ V > p_0 ] = e^{-\lambda p_0}
    \end{equation*}
    so 
    \begin{equation*}
        \Pr[V(1-r) > p_0 ] = e^{-\lambda \frac{p_0}{1-r}}.
    \end{equation*}
    
    Equation~\ref{eqn:expcondexp} follows from the memoryless property of exponential distributions, i.e.,
    
    \begin{align*}
        E\inbrak{V\suchthat V(1-r)>p_0} &= \frac{ \int_{\frac{p_0}{1-r}}^\infty x \lambda e^{-\lambda x} dx }{\Pr [(1-r)\cdot V > p_0] }\\
        &= e^{\lambda \frac{p_0}{1-r}} \int_{\frac{p_0}{1-r}}^\infty x \lambda e^{-\lambda x} dx \\
        &= e^{\lambda \frac{p_0}{1-r}} \inparen{ e^{-\lambda \frac{p_0}{1-r}} \cdot \frac{ \inparen{ \lambda \frac{p_0}{1-r} +1 }}{\lambda} } \\
        &= \frac{p_0}{1-r} + \frac{1}{\lambda} \\
        &= \frac{p_0}{1-r} + E[V].
    \end{align*}

    Plugging Equations~\ref{eqn:expprob} and \ref{eqn:expcondexp} into Equation~\ref{eqn:altuinfoasym},
    the creator revenue becomes
    \begin{equation}
        r \cdot \inparen{\frac{p_0}{1-r} + \frac{1}{\lambda}}\cdot e^{-\lambda \frac{p_0}{1-r}} + p_0e^{-\lambda p_0}.
    \end{equation}
    
    From here, setting $r=0$ gives the optimal revenues absent royalties, $p_0 e^{-\lambda p_0}$. Setting $p_0=0$ and $r=1$ gives the optimal revenues with royalties, $1/\lambda$. All other results follow trivially from here.
    
\end{proof}

\input{multi_unit_proofs}

\end{document}

%% file: multi_unit_proofs.tex
\subsection{Proofs for the Multi-Unit Setting \cref{sec:price_discrimination}}
\label{sec:multi-unit-proofs}

\input{multi_unit}

\begin{proof}[\bf Proof of \cref{lem:speculator-revenue}]

    When the speculator buys 0 units, it is easy to see that their expected revenue is 0, i.e., $\SR{0} = 0$.
    When the speculator buys 2 units, then they will sell them for prices $V_1$ and $V_2$, so their expected revenue is $2 \cdot E[V]$.

    The difficult case is when the speculator buys 1 unit, because in this case the speculator must compete with the creator.
    If the speculator purchases 1 unit, when it comes time to sell the unit to one of the end buyers, there are three possible scenarios:
    \begin{itemize}
        \item \textbf{Both below:} If $\max(v_1,v_2) < p_0$, then neither buyer is willing to buy from the creator (at price $p_0$) and the speculator sells their NFT to the high-value buyer for $\max(v_1,v_2) ~ \ord{V}{2}{2}$.
        \item \textbf{Split:} If $\min(v_1,v_2) < p_0 < \max(v_1,v_2)$, the speculator can still sell to the high-value buyer, but the speculator cannot set the price above $p_0$ (otherwise the high-value buyer would buy from the creator at $p_0$).  So in this case, the speculator sells their NFT for $p_0$
        \item \textbf{Both above:} If $p_0 < \min(v_1,v_2)$, then both buyers would be willing to buy from the creator.  In this case, we assume that the speculator (with better information) can identify the high-value buyer and sell to them (while the other buyer buys from the creator).  Thus in this case, the speculator sells their NFT for $\max(v_1,v_2) \sim \ord{V}{2}{2}$.
    \end{itemize}

    These three cases are illustrated in \cref{fig:multi-unit}.

    Thus the speculator can sell their unit to the end buyer for the price of
    \begin{align}
        \ord{V}{2}{2} &\mbox{ if $\ord{V}{2}{2} < p_0$} \\
        p_0 & \mbox{ if $\ord{V}{2}{1} < p_0 < \ord{V}{2}{2}$} \\
        \ord{V}{2}{2} &\mbox{ if $\ord{V}{2}{1} > p_0$}
    \end{align}

    Putting this together, the speculator's expected sale price is  
    \begin{equation}
        \label{eqn:speculator-revenue1a}
        E \inbrak{ \ord{V}{2}{2} \suchthat \ord{V}{2}{2} < p_0 } \cdot \Pr \inbrak{ \ord{V}{2}{2} < p_0 }
        + p_0 \cdot \Pr \inbrak{ \ord{V}{2}{1} < p_0 < \ord{V}{2}{2} }
        + E \inbrak{ \ord{V}{2}{2} \suchthat \ord{V}{2}{1} > p_0 } \cdot \Pr \inbrak{ \ord{V}{2}{1} > p_0 }
    \end{equation}

    See \cref{fig:multi-unit} for an illustration. 
    Now, $\Pr \inbrak{ \ord{V}{2}{2} < p_0 } = \inparen{ \Pr \inbrak{ V_1 < p_0 } \cdot \Pr \inbrak{ V_2 < p_0 } } = \inparen{F_V(p_0)}^2$, where $F_V(\cdot)$ is the CDF of $V$.  Similarly, $\Pr \inbrak{ \ord{V}{2}{1} > p_0 } = \Pr \inbrak{ V_1 > p_0 } \cdot \Pr \inbrak{ V_2 > p_0 } = \inparen{ 1 - F_V(p_0) }^2$. 

    This means that we can rewrite \cref{eqn:speculator-revenue1a} as
    \begin{align}
        \label{eqn:speculator-revenue1b}
        \mbox{Speculator expected revenue} &= 
        E \inbrak{ \ord{V}{2}{2} \suchthat \ord{V}{2}{2} < p_0 } \cdot \inparen{F_V(p_0)}^2 \\
        &+ p_0 \cdot \inparen{ 1 - \inparen{F_V(p_0)}^2 - \inparen{ 1 - F_V(p_0) }^2 } \\
        &+ E \inbrak{ \ord{V}{2}{2} \suchthat \ord{V}{2}{1} > p_0 } \cdot \inparen{ 1 - F_V(p_0) }^2 \\
        &= E \inbrak{ \ord{V}{2}{2} \suchthat \ord{V}{2}{2} < p_0 } \cdot \inparen{F_V(p_0)}^2 \\
        &+ 2 \cdot p_0 \cdot \inparen{ F_V(p_0) \cdot \inparen{ 1 - F_V(p_0) } } \\
        &+ E \inbrak{ \ord{V}{2}{2} \suchthat \ord{V}{2}{1} > p_0 } \cdot \inparen{ 1 - F_V(p_0) }^2
    \end{align}

    Now that $\SR{1}, \SR{2}, \SR{3}$ are well-defined, we can write the conditions under which the speculator would strictly prefer to buy $n$ over $\neg n$ units. These are simply obtained by comparing the relevant profits $(1-r)\SR{n}-p_0$, $n \in \{0,1,2\}$ for each case. As $(1-r)$ and $p_0$ will cancel out in these comparisons, the conditions simplify to: 
    \begin{align*}
        0 \mbox{ units } & \mbox{ if $\SR{0} > \max\inparen{ \SR{1}, \SR{2} }$} \quad \mbox{(high-price)}\\
        1 \mbox{ unit } \ & \mbox{ if $\SR{1} > \max\inparen{ \SR{0}, \SR{2} }$} \quad \mbox{(mid-price)}\\
        2 \mbox{ units } & \mbox{ if $\SR{2} > \max\inparen{ \SR{0}, \SR{1} }$} \quad \mbox{(low-price)}
    \end{align*}
    
    Next, notice that $\SR{1}$ explicitly depends on $p_0$, and thus the above conditions can be formally mapped to price levels, which we can informally refer to as $low-$, $mid-$ and $high-$price.
\end{proof}

\begin{proof}[\bf Proof of \cref{lem:creator-revenue-multi}]
    ~\\
    If the creator sells $k$ units to the speculator, then the creator receives $k \cdot p_0 + r \cdot \SR{k}$ from the speculator, 
    but the creator may also sell the remaining units to the end buyers.  So we calculate the creator's expected revenue as follows:\\
    \textbf{Speculator buys 0 units:}\\
        In this case, the creator retains both units, and their expected revenue is 
        \begin{equation}\label{eqn:creator-revenue-high}
            \CR{0} = 2 \cdot p_0 \cdot \Pr \inbrak{ V > p_0 } = 2 \cdot p_0 \cdot \inparen{ 1 - F_V(p_0) }
        \end{equation} 
    \vspace{1em}
    \textbf{Speculator buys 1 unit:}\\
        In this case, the creator retains one unit, and the speculator buys one unit. The creator's expected revenue is 
        \begin{equation}\label{eqn:creator-revenue-mid}
            \CR{1} = \underbrace{p_0}_{\mbox{Payment from speculator}} + \underbrace{r \cdot \SR{1}}_{\mbox{Royalties from speculator}} + \underbrace{p_0 \cdot \Pr \inbrak{ \ord{V}{1}{2} > p_0 }}_{\mbox{Payment from end-buyers}}
        \end{equation} 
        where $\SR{1}$ is the speculator's expected revenue when they buy 1 unit (given in \cref{eqn:speculator-revenue1}).
        Note that because the speculator always sells to the higher valuation buyer, the only way the creator can sell the second unit is if the 
        \emph{lower-valued} end-buyer is willing to pay $p_0$, i.e., if $\ord{V}{1}{2} > p_0$.\\
        \vspace{1em}
    \textbf{Speculator buys 2 units:}\\
        In this case, the creator retains both units, and the speculator buys both units. The creator's expected revenue is 
        \begin{equation}\label{eqn:creator-revenue-low}
            \CR{2} = 2 \cdot \inparen{ p_0 + r \cdot E[V] }
        \end{equation}

\end{proof}

\begin{figure}[ht]
    \begin{center}
        \begin{tikzpicture}
            \pgfmathsetmacro{\t}{2}
            \draw [->,thick] (0,0) -- (4,0); 
            \draw [->,thick] (0,0) -- (0,4); 

            \draw [dashed] (\t,-.1) -- (\t,4);
            \draw [dashed] (-.1,\t) -- (4,\t);
           
            \pgfmathsetmacro{\tt}{\t/2}
            \pgfmathsetmacro{\ttt}{3*\t/2}
            \node at (\tt,\tt) {$A$};
            \node at (\ttt,\ttt) {$B$};
            \node at (\tt,\ttt) {$C$};
            \node at (\ttt,\tt) {$D$};

            \node [anchor=north]at (\t,-.2) {$p_0$};
            \node [anchor=east]at (-.2,\t) {$p_0$};

        \end{tikzpicture}
        \caption{Buyer valuations in the two-unit case.  In region $A$, both buyers have valuations less than $p_0$.
        In region $B$, both buyers have valuations \emph{greater} than $p_0$.  In regions $C$ and $D$, one buyer has a valuation
        less than $p_0$ and the other buyer has a valuation greater than $p_0$. \label{fig:two-unit-valuation}}
    \end{center}
\end{figure}

By partitioning the space into $A$, $B$, $C$, and $D$, we can more easily compute the speculator's expected revenue when they buy one unit, $\SR{1}$.
Using this partition, we can rewrite the terms of \cref{eqn:speculator-revenue1b} as follows:
\begin{align}
    E \inbrak{ \ord{V}{2}{2} \suchthat \ord{V}{2}{2} < p_0 } \cdot \Pr \inbrak{ \ord{V}{2}{2} < p_0 } = \int_A \max(x,y) f_V(x) f_V(y) dx dy \\
    E \inbrak{ \ord{V}{2}{2} \suchthat \ord{V}{1}{2} > p_0 } \cdot \Pr \inbrak{ \ord{V}{1}{2} > p_0 } = \int_B \max(x,y) f_V(x) f_V(y) dx dy \\
    2 \cdot p_0 \cdot \inparen{ 1 - \Pr \inbrak{ \ord{V}{2}{2} < p_0 } - \Pr \inbrak{ \ord{V}{1}{2} > p_0 } } = \int_{C \cup D} p_0 \cdot f_V(x) f_V(y) dx dy
\end{align}

If we assume that the speculator buys one unit, then the speculator pays $p_0$ for 
that one item, and so in region $A$, the speculator \emph{loses} money, 
while in region $B$ the speculator makes money and in regions $C$ and $D$ the speculator breaks even.

Below, we provide 4 technical lemmas that will be useful for the remaining proofs.
\begin{lemma}
    \label{lem:total-revenue}
    If $V$ is non-degenerate, then the total revenue to both the buyer and speculator is maximized 
    when the speculator buys two units.

    In particular

    \begin{align}
        (1-r) \cdot \SR{2} - 2p_0 + \CR{2} = 2 \cdot E[V] \qquad \mbox{(two units)} \\
        (1-r) \cdot \SR{1} - p_0 + \CR{1} < 2 \cdot E[V] \qquad \mbox{(one unit)} \\
        (1-r) \cdot \SR{0} + \CR{0} < 2 \cdot E[V] \qquad \mbox{(zero units)}
    \end{align}
\end{lemma}

\begin{proof}[\bf Proof of \cref{lem:total-revenue}]
    For the two unit case, note that 
    \begin{align}
        (1-r) \cdot \SR{2} -2p_0 + \CR{2} &= 2 \cdot p_0 + r \cdot \SR{2} \\
        &= \SR{2} \\
        &= 2 \cdot E[V]
    \end{align}

    Next, for the one unit case, note that

    \begin{align*}
        (1-r) \cdot \SR{1} - p_0 + \CR{1} &= (1-r) \cdot \SR{1} - p_0 + p_0 + r \cdot \SR{1} + p_0 \cdot \Pr \inbrak{ \ord{V}{1}{2} > p_0 } \\
        &= \SR{1} + p_0 \cdot \Pr \inbrak{ \ord{V}{1}{2} > p_0 } \\
        &=  E \inbrak{ \ord{V}{2}{2} \suchthat \ord{V}{2}{2} < p_0 } \cdot \inparen{F_V(p_0)}^2 \qquad \mbox{(from \cref{eqn:speculator-revenue1b})} \\
        &+ 2 \cdot p_0 \cdot \inparen{ F_V(p_0) \cdot \inparen{ 1 - F_V(p_0) } } \\
        &+ E \inbrak{ \ord{V}{2}{2} \suchthat \ord{V}{1}{2} > p_0 } \cdot \inparen{ 1 - F_V(p_0) }^2 \\
        &+ p_0 \cdot \inparen{ 1 - F_V(p_0) }^2 \\
        &=  E \inbrak{ \ord{V}{2}{2} \suchthat \ord{V}{2}{2} < p_0 } \cdot \inparen{F_V(p_0)}^2 \qquad \mbox{(from \cref{eqn:speculator-revenue1b})} \\
        &+ p_0 \cdot \inparen{ 1 - \inparen{F_V(p_0)}^2 } \\
        &+ E \inbrak{ \ord{V}{2}{2} \suchthat \ord{V}{1}{2}> p_0 } \cdot \inparen{ 1 - F_V(p_0) }^2 \\
        &= \int_{A \cup B} \max(x,y) f_V(x) f_V(y) dx dy \\
        &+ \int_{B \cup C \cup D} p_0 \cdot f_V(x) f_V(y) dx dy  \\
        &= \int_{A} \max(x,y) f_V(x) f_V(y) dx dy \\
        &+ \int_{B} \inparen{ \max(x,y) + p_0 } f_V(x) f_V(y) dx dy + \int_{C \cup D} p_0 \cdot f_V(x) f_V(y) dx dy  \\
        &\le \int_{A} (x+y) f_V(x) f_V(y) dx dy + \int_{B} ( \max(x,y) + p_0 ) f_V(x) f_V(y) dx dy \\ 
        &+ \int_{C \cup D} p_0 \cdot f_V(x) f_V(y) dx dy \\
        &< \int_{A} (x+y) f_V(x) f_V(y) dx dy \\
        &+ \int_{B} ( \max(x,y) + \min(x,y) ) f_V(x) f_V(y) dx dy \\
        &+ \int_{C \cup D} p_0 \cdot f_V(x) f_V(y) dx dy \\
        &= \int_{A} (x+y) f_V(x) f_V(y) dx dy + \int_{B} ( x+y) f_V(x) f_V(y) dx dy\\
        &(...)
    \end{align*}
    \begin{align*}
    &(...)\\
    &+ \int_{C \cup D} p_0 \cdot f_V(x) f_V(y) dx dy \\
        &< \int_{A} (x+y) f_V(x) f_V(y) dx dy + \int_{B} ( x+y) f_V(x) f_V(y) dx dy \\
        &+ \int_{C \cup D} (x+y) \cdot f_V(x) f_V(y) dx dy \\
        &= \int_{A \cup B \cup C \cup D} (x+y) f_V(x) f_V(y) dx dy \\
        &= 2E[V]
    \end{align*}

    So $(1-r) \cdot \SR{1} - p_0 + \CR{1} < 2 E[V]$.

    Finally, for the zero unit case, we have $\SR{0} = 0$, so $\SR{0} + \CR{0} = \CR{0}$, and we 
    just need to show that $\CR{0} < 2 \cdot E[V]$.

    From \cref{eqn:creator-revenue-high}, we have
    \begin{align}
        \CR{0} &= 2 \cdot p_0 \cdot \inparen{ 1 - F_V(p_0) }
    \end{align}

    Now, if $p_0 < E[V]$, then 
    \begin{equation}
       \CR{0} = 2 \cdot p_0 \cdot \inparen{ 1 - F_V(p_0) } < 2 \cdot E[V] \cdot \inparen{ 1 - F_V(E[V]) } \le 2 \cdot E[V]
    \end{equation} 
    so the creator cannot extract the maximum revenue.

    On the other hand, if $p_0 = E[V]$, then because $V$ is non-degenerate, we have $F_V(E[V]) < 1$.
    Thus 
    \begin{align}
        \CR{0} &= 2 \cdot p_0 \cdot \inparen{ 1 - F_V(p_0) } = 2 \cdot E[V] \cdot \inparen{ 1 - F_V(E[V]) } < 2 \cdot E[V]
    \end{align}

    Finally, if $p_0 > E[V]$, then 
    \begin{align}
        E[V] &= \int_0^\infty x f_V(x) dx \\
        &= \int_0^{p_0} x f_V(x) dx + \int_{p_0}^\infty x f_V(x) dx \\
        &\ge \int_0^{p_0} x f_V(x) dx + p_0 \cdot \inparen{ 1 - F_V(p_0) }
    \end{align}
    But since $p_0 > E[V]$ and $V$ is non-degenerate, we must have $\int_0^{p_0} x f_V(x) dx > 0$.
    So 
    \begin{align}
        \CR{0} &= 2 \cdot p_0 \cdot \inparen{ 1 - F_V(p_0) } < 2 \cdot E[V]
    \end{align}
    so the creator cannot extract the maximum revenue.
\end{proof}

\begin{lemma}
    \label{lem:multi-unit-no-royalties-suboptimal}

    In the two unit setting, the only way the creator can extract the maximum revenue of $2 \cdot E[V]$ is if the speculator buys two units 
    and either $r = 0$ and $p_0 = E[V]$, or $r = 1$ and $p_0 = 0$.
\end{lemma}

\begin{proof}[\bf Proof of \cref{lem:multi-unit-no-royalties-suboptimal}]
    From \cref{lem:total-revenue}, we have that $\CR{0} < 2 \cdot E[V]$ and $\CR{1} < 2 \cdot E[V]$, 
    so the only way the creator can extract the maximum possible revenue ($2 \cdot E[V]$) is if the speculator buys two units.

    If the speculator buys two units, the end buyers pay a total of $2 \cdot E[V]$ and this is split between the creator and speculator.
    In this case, the only way the creator can extract the maximum revenue is if the speculator makes zero profit, i.e.,
    
    \begin{align}
        (1-r) \cdot \SR{2} - 2p_0 = (1-r) \cdot 2 \cdot E[V] - 2p_0 = 0
    \end{align}

    This implies either $r = 0$ and $p_0 = E[V]$, or $r = 1$ and $p_0 = 0$.
\end{proof}

\begin{lemma}
    \label{lem:speculator-revenue-multi-exponential}
    When there are two units, and the buyers' valuations are exponentially distributed with parameter $\lambda$, 
    then,
    \begin{align}
        \SR{0} &= 0 \\
        \SR{1} &= \frac{1}{2\lambda} \cdot \inparen{ 3 + 4 e^{-2p_0 \lambda} - 4e^{-p_0 \lambda}} \label{eqn:speculator-revenue1-exp} \\
        \SR{2} &= \frac{2}{\lambda} \label{eqn:speculator-revenue2-exp}
    \end{align}
\end{lemma}

\begin{proof}[\bf Proof of \cref{lem:speculator-revenue-multi-exponential}]
    For any distribution, $\SR{0} = 0$, since the speculator cannot make a profit if they buy 0 units.

    For the exponential distribution, we have
    \begin{align}
        E[V] &= \frac{1}{\lambda} \\
        F_V(x) &= 1 - e^{-\lambda x} \\
        \Pr \inbrak{ \ord{V}{2}{2} < t } &= F_V(t)^2 = \inparen{ 1 - e^{-\lambda t} }^2 \\
        \Pr \inbrak{ \ord{V}{1}{2} > t } &= \inparen{ 1 - F_V(t) }^2 = e^{-2\lambda t} \\
    \end{align}

    First, we compute the speculator expected revenue \emph{if they buy one unit}.
    The general form for $\SR{1}$ is given in \cref{eqn:speculator-revenue1b}.
    So we need to calculate the terms in \cref{eqn:speculator-revenue1b}.

    \begin{align}
        E \inbrak{ \ord{V}{2}{2} \suchthat \ord{V}{2}{2} < t } \cdot \inparen{ F_V(t) }^2 &= \int_0^t \int_y^t x f_V(x) f_V(y) dx dy \\
        &= 2 \int_0^t \int_y^t x e^{-\lambda x} e^{-\lambda y} dx dy \\
        &= 2\int_0^t e^{-\lambda y} \int_y^t x e^{-\lambda x} dx dy \\
        &= 2 \int_0^t e^{-\lambda y} \inbrak{ \frac{1}{\lambda} \inparen{ e^{-\lambda(t+y)} \inparen{ e^{t\lambda} (1 + y\lambda) - e^{y\lambda} (1 + t\lambda) }}} dy \\
        &= \frac{1}{2\lambda} \cdot \inparen{ e^{-2t \lambda} \inparen{ 1 + 3e^{2t\lambda} + 2t\lambda -4e^{t\lambda}(1+t\lambda)}} 
    \end{align}

    \begin{align}
        E \inbrak{ \ord{V}{2}{2} \suchthat \ord{V}{1}{2} > t } \cdot \inparen{ 1 - F_V(t) }^2 &= 2\int_t^\infty \int_y^\infty x f_V(x) f_V(y) dx dy \\
        &= 2\int_t^\infty f_V(y) \int_y^\infty x f_V(x) dx dy \\
        &= 2\int_t^\infty f_V(y) \inbrak{ \frac{e^{-2t\lambda}(3+2t\lambda)}{4\lambda} }dy \\
        &= \frac{1}{2\lambda} \cdot \inparen{ e^{-2t\lambda}(3+2t\lambda) }
    \end{align}

    Putting this together, we have
    \begin{align}
        \SR{1} &= \frac{1}{2\lambda} \cdot \inparen{ e^{-2p_0 \lambda} \inparen{ 1 + 3e^{2p_0 \lambda} + 2p_0 \lambda -4e^{p_0 \lambda}(1+p_0 \lambda)}} \\
        &+ 2 \cdot p_0 \cdot e^{-p_0 \lambda} \cdot \inparen{ 1 - e^{-p_0 \lambda} } \\
        &+ \frac{1}{2\lambda} \cdot \inparen{ e^{-2p_0 \lambda}(3+2p_0 \lambda) }\\
        &= \frac{1}{2\lambda} \cdot \inparen{ 3 + 4 e^{-2p_0 \lambda} - 4e^{-p_0 \lambda}}
    \end{align}

    Finally, $\SR{2} = 2 \cdot E[V] = \frac{2}{\lambda}$.

\end{proof}

\begin{lemma}[Creator Revenue in the multi-unit setting with exponential valuations {c.f. \cref{lem:creator-revenue-multi}}]
    \label{lem:creator-revenue-multi-exponential}
    When there are two units, and the buyers' valuations are exponentially distributed with parameter $\lambda$, 
    then,
    \begin{align}
        \CR{0} &= 2 \cdot p_0 \cdot e^{-\lambda p_0} \label{eqn:creator-rev-exp-0} \\
        \CR{1} &= p_0 \inparen{ 1 + e^{-2p_0 \lambda } }+ r \cdot \frac{1}{2\lambda} \cdot \inparen{ 3 + 4 e^{-2p_0 \lambda} - 4e^{-p_0 \lambda}} \label{eqn:creator-rev-exp-1} \\
        \CR{2} &= 2 \cdot \inparen{ p_0 + r \cdot \frac{1}{\lambda} } \label{eqn:creator-rev-exp-2}
    \end{align}
\end{lemma}

\begin{proof}[\bf Proof of \cref{lem:creator-revenue-multi-exponential}]
    From \cref{eqn:creator-revenue-high}, we have
    \begin{align}
        \CR{0} &= 2 \cdot p_0 (1 - F_V(p_0)) = 2 \cdot p_0 \cdot e^{-\lambda p_0}
    \end{align}

    From \cref{eqn:creator-revenue-mid}, we have
    \begin{align}
        \CR{1} &= p_0 + r \cdot \SR{1} + p_0 \cdot \Pr \inbrak{ \ord{V}{1}{2} > p_0 } \\
                &= p_0 + r \cdot \frac{1}{2\lambda} \cdot \inparen{ 3 + 4 e^{-2p_0 \lambda} - 4e^{-p_0 \lambda}} + p_0 \cdot e^{-2p_0 \lambda} \\
                &= p_0 \inparen{ 1 + e^{-2p_0 \lambda } }+ r \cdot \frac{1}{2\lambda} \cdot \inparen{ 3 + 4 e^{-2p_0 \lambda} - 4e^{-p_0 \lambda}}
    \end{align}

    From \cref{eqn:creator-revenue-low}, we have
    \begin{align}
        \CR{2} &= 2 \cdot \inparen{ p_0 + r \cdot E[V] } \\
                &= 2 \cdot \inparen{ p_0 + r \cdot \frac{1}{\lambda} }
    \end{align}
\end{proof}

\begin{proof}[\bf Proof of \cref{thm:multi-unitr0}]

    We will show a stronger result:
    in the two unit setting, if $V$ is exponentially distributed with parameter $\lambda$, 
    and there are \emph{no royalties}, then the creator's revenue is at most $.971 \cdot 2 \cdot E[V]$.

    The speculator will purchase two units (instead of one unit) if
    \begin{align}
        2 (1-r) \cdot \SR{2} - 2p_0 &\ge (1-r) \SR{1} - p_0 \\
        &\Updownarrow \nonumber \\
        \frac{2 (1-r)}{\lambda} - 2p_0 &\ge (1-r) \SR{1} - p_0 \\
        &\Updownarrow \nonumber \\
        \frac{2 (1-r)}{\lambda} - 2p_0 &\ge \frac{1-r}{2\lambda} \cdot \inparen{ 3 + 4 e^{-2p_0 \lambda} - 4e^{-p_0 \lambda}} - p_0 \\
        &\Updownarrow \nonumber \\
        \frac{1-r}{2\lambda} \inparen{ 1 + 4 e^{-p_0 \lambda} - 4e^{-2p_0 \lambda}} &\ge p_0 \\
        &\Updownarrow \nonumber \\
        (1-r) \inparen{ 1 + 4 e^{-p_0 \lambda} - 4e^{-2p_0 \lambda}} &\ge 2 \cdot \lambda \cdot p_0 \label{eqn:speculator-purchase-two-units}
    \end{align}

    Making the substition $x = p_0 \cdot \lambda$, we have the condition
    \begin{equation}
        (1-r) \inparen{ 1 + 4 e^{-x} - 4e^{-2x}} \ge 2 \cdot x
    \end{equation}
    It's not hard to check that $1 + 4 e^{-x} - 4e^{-2x} \le 2$ (with the maximum occurring at $x = \ln(2)$).
    We can also see that if $x \ge .971$, then the inequality in \cref{eqn:speculator-purchase-two-units} is violated, 
    so the speculator prefers to purchase 
    one unit instead of two.

    If $p_0 \cdot \lambda < .971$, then
    \begin{equation}
        \CR{2} = 2 \cdot p_0  <  .971 \cdot \frac{2}{\lambda} = .971 \cdot 2 \cdot E[V]
    \end{equation}

    Now, we just need to check that the creator cannot make more than $.971 \cdot 2 \cdot E[V]$ if the speculator buys 0 or 1 unit.
    From \cref{eqn:creator-rev-exp-0}, we have $\CR{0} = 2 \cdot p_0 \cdot e^{-\lambda p_0} = \frac{1}{\lambda} \cdot 2 \cdot x \cdot e^{-x}$.
    This is maximized at $x = 1$, and in this case $\CR{0} = \frac{2}{e\lambda} < .971 \cdot 2 \cdot E[V]$.
    So the creator cannot make more than $.971 \cdot 2 \cdot E[V]$ by selling 0 units.

    Similarly, from \cref{eqn:creator-rev-exp-1}, we have
    \begin{equation}
        \CR{1} = p_0 \inparen{ 1 + e^{-2p_0 \lambda } }+ r \cdot \frac{1}{2\lambda} \cdot \inparen{ 3 + 4 e^{-2p_0 \lambda} - 4e^{-p_0 \lambda}}
    \end{equation}
   
    Using the above substition, $x = p_0 \cdot \lambda$, then
    \begin{equation}
        \CR{1} = \frac{1}{ \lambda} \inparen{ x \inparen{ 1 + e^{-2x} }+ r \cdot \frac{1}{2} \cdot \inparen{ 3 + 4 e^{-2x} - 4e^{-x}} }
    \end{equation}
    When $r = 0$, this becomes
    \begin{equation}
        \CR{1} = \frac{1}{ \lambda} \inparen{ x \inparen{ 1 + e^{-2x} } }
    \end{equation}
    which is strictly increasing in $x$ (and thus in $p_0$).

    From \cref{eqn:speculator-revenue1-exp}, we have that 
    \begin{equation}
        \SR{1} = \frac{1}{2\lambda} \cdot \inparen{ 3 + 4 e^{-2x} - 4e^{-x}}
    \end{equation}

    The speculator revenue is $(1-r) \cdot \SR{1} - p_0$, which is 
    \begin{equation}
        \frac{1}{\lambda} \inparen{ \frac{1-r}{2} \cdot \inparen{ 3 + 4 e^{-2x} - 4e^{-x}} - x }
    \end{equation}
    This is strictly negative for $x \ge 1.05$.

    Thus if the creator sets $p_0 > \frac{1.05}{\lambda}$, then the speculator will prefer to purchase 0 units, 
    so the maximum revenue the creator can get in the one unit case is 
    \begin{equation}
        \CR{1} = \frac{1}{\lambda} \inparen{ x \inparen{ 1 + e^{-2x} } } \le \frac{1}{\lambda} \cdot \inparen{ 1.05 + e^{-2.1} } \sim \frac{1.17}{\lambda} < .971 \cdot \frac{2}{\lambda}
    \end{equation}
    which is strictly decreasing in $x$ (and thus in $p_0$).

    So the creator cannot make more than $.971 \cdot 2 \cdot E[V]$ by selling 0 or 1 units.

    To understand why this happens, it's helpful to look at some of our previous results visually. Figures~\ref{fig:2unit_revenue_exp} and \ref{fig:2unit_regions_exp} show creator revenues and speculator profits, respectively, as a function of the chosen $p_0$. The region over which the speculator prefers to buy 1 unit is rather small (it can become larger under alternative distributional assumptions), but it illustrates the speculator's optionality and why royalties can still add value in this case, even in the absence of information asymmetry.

\begin{figure}[H]
    \centering
    \begin{subfigure}[b]{0.48\textwidth}
        \includegraphics[width=\textwidth]{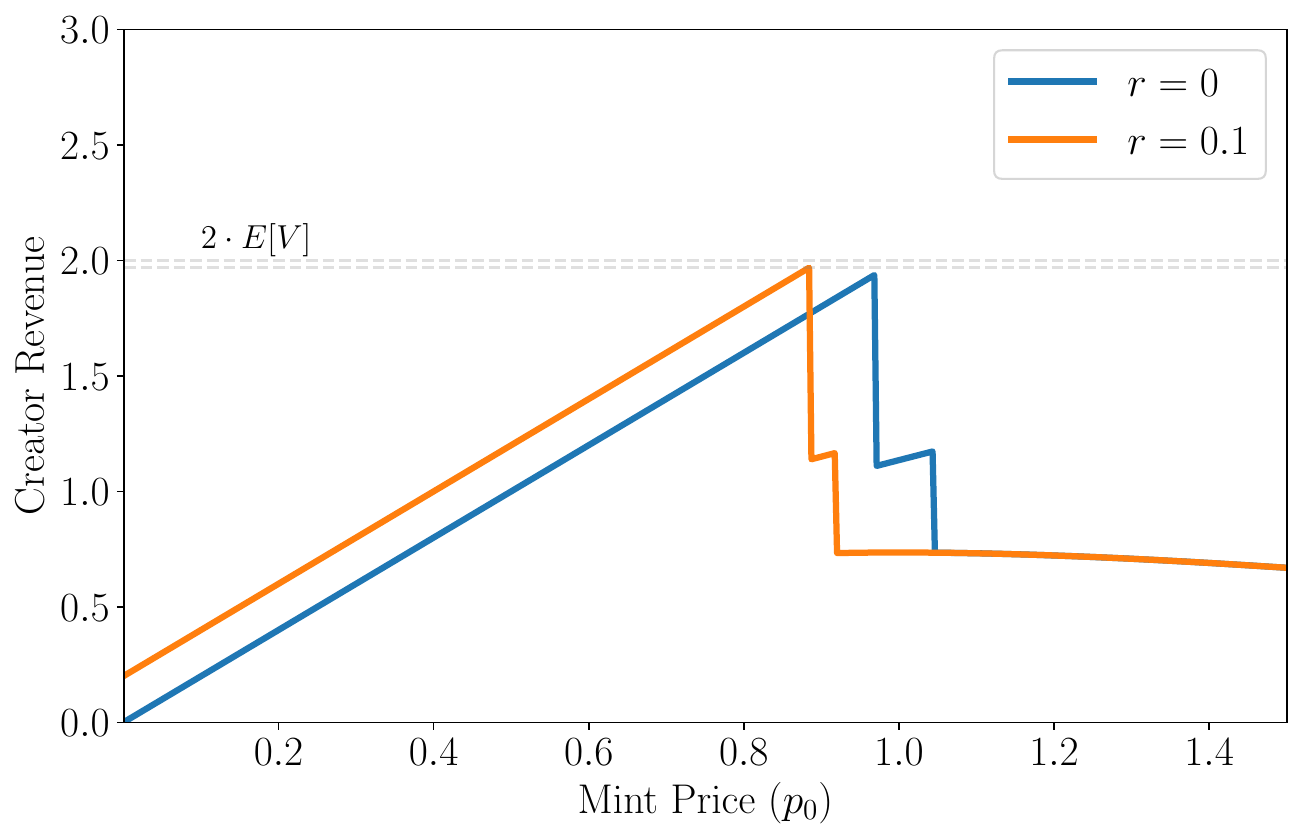}
        \caption{Creator revenue}
        \label{fig:2unit_revenue_exp}
    \end{subfigure}
    \hfill 
    \begin{subfigure}[b]{0.48\textwidth}
        \includegraphics[width=\textwidth]{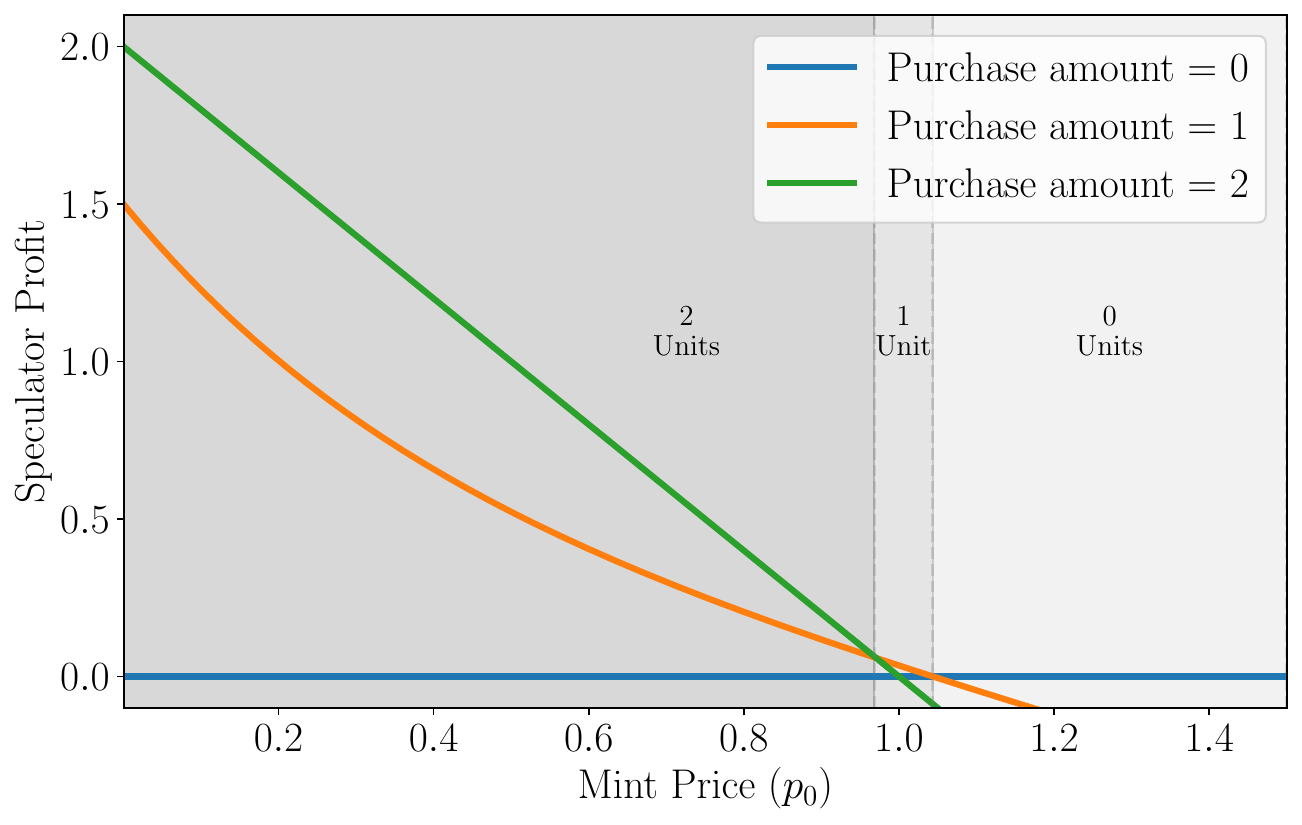} 
        \caption{Speculator's purchase decision}
        \label{fig:2unit_regions_exp}
    \end{subfigure}
    \caption{Creator revenues from \cref{lem:creator-revenue-multi} (with $\lambda=1$) as a function of mint price $p_0$, when the creator has 2 NFTs to sell. The optimal mint price in each case corresponds to the peak point.}
\end{figure}

\end{proof}

\begin{proof}[\bf Proof of \cref{lem:multi_exponential}]
    From the proof of \cref{thm:multi-unitr0}, we have that the creator's revenue without royalties is at most $.971 \cdot 2 \cdot E[V]$.
    From \cref{lem:creator-revenue-multi-exponential}, we have that the creator's revenue with royalties is $2 \cdot E[V]$.
    So the creator's revenue with royalties is at least $2 \cdot E[V] - .971 \cdot 2 \cdot E[V] = .029 \cdot 2 \cdot E[V]$.
\end{proof}

%% file: multi_unit.tex
\begin{figure}[H]
    \centering
    \begin{tikzpicture}[scale=3]
        \node [inner sep=0] {} [grow'=right]
            child {node {$\ord{V}{2}{2}$} 
                edge from parent node [above,anchor=south, rotate=30] {$\ord{V}{2}{2} < p_0$}}
            child {node [inner sep=0] {}
                child {node {$p_0$}
                edge from parent node [above,anchor=south,rotate=30] {$\ord{V}{1}{2} < p_0$} }
                child {node {$\ord{V}{2}{2}$}
                edge from parent node [below,anchor=north,rotate=-30] {$\ord{V}{1}{2} > p_0$} }
                edge from parent node [below,anchor=north,rotate=-30] {$\ord{V}{2}{2} > p_0$}
                };
    \end{tikzpicture}
    \caption{Speculator revenue when speculator purchases 1 (out of the 2) units.}
    \label{fig:multi-unit}
\end{figure}